\documentclass{scrartcl}       
\usepackage{graphicx}
\usepackage[utf8]{inputenc}					
\usepackage[T1]{fontenc}					
\usepackage{ae,aecompl}         			
\usepackage[english]{babel}         		
\usepackage{amsmath,amssymb,amstext,amsbsy}	
\usepackage{exscale}                		
\usepackage{calc}
\usepackage{subfigure}%
\usepackage{lineno}
\usepackage{authblk}
\usepackage[dvipsnames]{xcolor}
\usepackage{matlab-prettifier}
\usepackage{fixmath}
\usepackage{multirow}
\usepackage[round]{natbib}

\usepackage{color}
\definecolor{light}{gray}{0.50}
\definecolor{heavy}{gray}{0.35}
\definecolor{black}{gray}{0.0}
\definecolor{dgreen}{rgb}{0.0,0.7,0}
\definecolor{dred}{rgb}{0.9959,0,0}
\definecolor{green}{rgb}{0.0,0.99599,0.0}
\definecolor{purple}{rgb}{0.6,0.0,0.4}

\newcounter{kleincommentno}
\newcounter{puetzcommentno}
\usepackage{array}
\usepackage{marvosym}

\usepackage{tikz}
\newcommand\diag[4]{%
  \multicolumn{1}{p{#2}|}{\hskip-\tabcolsep
  $\vcenter{\begin{tikzpicture}[baseline=0,anchor=south west,inner sep=#1]
  \path[use as bounding box] (0,0) rectangle (#2+2\tabcolsep,\baselineskip);
  \node[minimum width={#2+2\tabcolsep},minimum height=\baselineskip+\extrarowheight] (box) {};
  \draw (box.north west) -- (box.south east);
  \node[anchor=south west] at (box.south west) {#3};
  \node[anchor=north east] at (box.north east) {#4};
 \end{tikzpicture}}$\hskip-\tabcolsep}}

\usepackage[
    colorlinks=true,
    urlcolor=black,
    citecolor=blue,
    linkcolor=blue
  ]{hyperref}

\newcommand{\R}{\mathbb{R}}

\newcommand{\wdef}{\mathrel{\mathop{:}}=}
\newcommand{\wdefi}{=\mathrel{\mathop{:}}}
\newcommand{\labs}{\left|}
\newcommand{\rabs}{\right|}

\newcommand{\pd}[2]{{\frac{\partial  #1}{\partial  #2}}}
\newcommand{\td}[2]{{\frac{\operatorname{d}\!  #1}{\operatorname{d}\!  #2}}}

\usepackage{authblk}

\begin{document}

\title{Reflection and transmission of gravity waves at non-uniform stratification layers
}
%
%
%



\author[1]{Christopher Pütz}
\author[1]{Mark Schlutow}
\author[1]{Rupert Klein}
\author[2]{Vera Bense}
\author[2]{Peter Spichtinger}

\affil[1]{Department of Mathematics, Freie Universität Berlin, Germany}
\affil[2]{Institute for Atmospheric Physics, Johannes Gutenberg Universität Mainz, Germany}
\date{}



\maketitle

\begin{center}
\textsc{Abstract}
\end{center}
{\small
The present study focuses on the interaction of gravity waves in the atmosphere with the tropopause. As the vertical extent of the latter is small compared to the density scale height, wave propagation is described by the Taylor-Goldstein equation as derived from the linearised Boussinesq approximation. Of particular interest in the construction of gravity wave parameterisations for the upper atmosphere are the transmission and reflection properties of the tropopause as these determine the upward fluxes of energy and momentum carried by internal waves. 

A method is presented that decomposes internal waves explicitly into upward and downward propagating contributions, thus giving direct access to transmission and reflection coefficients of finite regions of non-uniform stratification in a stationary atmosphere. The scheme utilizes a piecewise constant approximation for the background stratification and matches up- and downward propagating plane wave solutions in each layer through physically meaningful coupling conditions. As a result, transmission and reflection coefficients follow immediately. 

In the limit of an increasing number of layers the method leads to a reformulation of the Taylor-Goldstein equation in a particular set of variables. Numerical integration of this non-constant coefficient differential equation provides a representation of Taylor-Goldstein solutions that also distinguishes explicitly between the upward and downward travelling wave branches of the dispersion relation and hence gives access to transmission coefficients also for smoothly stratified layers. 

The multi-layer solutions are not only shown to converge to the limit solution quadratically with the number of layers, but are also found to be surprisingly accurate -- and hence efficient -- for very small numbers of vertical layers. The results obtained for some test cases are in good agreement with several existing results as well as with two-dimensional numerical solutions of the full non-linear pseudo-incompressible equations for a vertical slice. Yet, by revealing the up- and downward travelling wave components explicitly, the multi-layer solutions offer alternative insights into the interaction of gravity waves propagating through non-uniform stratification. 

The present paper focuses on internal wave eigenmodes of a stratified atmosphere. Yet, it also serves as the basis for the development of a new numerical method for the propagation of non-stationary wave packets described in a companion paper. 


}
\newpage

\section{Introduction}

Gravity waves arise from fluid displacements in a vertically stably stratified medium, in which the buoyancy acts as restoring force. This includes, to a large part, also the earth's atmosphere. The propagation of atmospheric gravity waves has been the subject of a number of earlier studies. One of the first investigations goes back to \citet{Scorer49}, who focused on orographically generated waves and how they can be trapped in the lee of a mountain ridge. 

An important characteristic of gravity waves is the ability to transport energy horizontally as well as vertically. \citet{EliassenPalm61} analysed waves carrying energy upward and downward in the context of orographically excited gravity waves that reflect from vertically varying stratification and wind. They used a piecewise-constant approximation for both stratification and wind and found local solutions. These were matched at the discontinuities of the approximation. The calculations were carried out for a two- and a three-layer atmosphere. The multi-layer method to be introduced below is also based on these concepts.
 
By a similar approach \citet{DanielsenBleck70} examined mountain waves and approximated key atmospheric parameters by piecewise exponential functions, which allowed them to solve the governing equations by combinations of Bessel functions. \citet{SutherlandYewchuck04} attended to the topic again and scrutinised the phenomenon of wave tunnelling, which describes the energy transport over a finite layer of decreased, or even vanishing, stratification. They undertook a mathematical analysis as well as laboratory experiments to support their findings. \citet{BrownSutherland06} expanded the theory by allowing for shear flow over an unstratified layer. Both scenarios were later examined numerically by \citet{NaultSutherland07} who provided numerical solutions for plane wave transmission in arbitrary stratification and wind. As a concrete example, they performed simulations for an atmospheric stratification and wind profile that was observed over Jan Mayen island. 

Diffraction through a slit and back-reflection from a slope is covered by \citet{MercierETAL08}, mainly through experimental work. To keep track of the direction of the wave propagation, they use a demodulation of the measured wave signal. \citet{Buehler09} uses a method similar to that introduced by \citet{EliassenPalm61} for the two-dimensional shallow water equations with water depth that changes only with one of the spatial coordinates. It is approximated by a piecewise-constant depth which allows for explicit solutions in each of those layers.

The main focus of the present work is on the transmission and reflection of gravity waves at confined regions of non-uniform stratification. The governing equations are the linearised Boussinesq equations, which can be combined into a single equation for one of the dynamic variables. Since this ``Taylor-Goldstein equation'' does not allow for explicit solutions in general, we use a piecewise-constant approximation of the stratification with finitely many layers to generate an analytically accessible approximate solution from which we can compute the transmission and reflection of gravity waves. In each layer of constant stratification an explicit solution is constructed as a superposition of upward- and downward-travelling plane waves. These are matched at the discontinuities of the piecewise-constant approximation of the background rendering the perturbations of vertical wind and pressure continuous across these interfaces. Conservation of energy is used to derive a transmission coefficient. 

By a reformulation, we find a new set of differential equations that describes the limit of an infinite number of discrete layers. These limit equations are solved numerically. The resulting transmission coefficients are then compared with those calculated for a finite number of layers. For an increasing number of layers, the solution is found to converge to the limit solutions with second order accuracy. Yet, even for a moderate number of layers the results turn out to be rather accurate, and this allows for very efficient computational estimates of the transmission and reflection properties of tropopause-like layers in the atmosphere. Another advantage of the presented method is its ability to keep track of upward and downward propagating waves \emph{locally} in arbitrary stratification profiles. Moreover, an extension of this method to a solver for non-stationary wave packets has been achieved, and this is the topic of the companion paper \citep{PuetzKlein2018}.

The present paper is structured as follows. In section~\ref{sec:MultiLayerMethod} we present the method developed to analytically compute a transmission coefficient. Numerical computations are pursued in section~\ref{sec:convergence} to provide evidence of the correctness of the method as well as an error analysis. We show results for selected stratification profiles in section~\ref{sec:results}. Section~\ref{sec:numerics} summarizes results from numerical solutions of the full Boussinesq equations to further back up our theoretical findings. 
In section~\ref{sec:FurtherDiscussion}, we discuss further extensions as well as limitations of the method. In particular, the adaptation to shear layers and wave packets is of special interest for further studies.


\section{Multi-layer method}
\label{sec:MultiLayerMethod}

We are interested in wave propagation through the tropopause. The latter is characterised by sharp changes in the stratification and also by strong jet winds. Above and below the tropopause, we have almost uniform stratification. The extent of the tropopause is small compared to the density scale height. Therefore, as we only focus on the propagation through this rather shallow area, we can apply the Boussinesq approximation.

The starting points for our investigation are the two-dimensional inviscid Boussinesq equations with small amplitudes and an atmosphere at rest.
If we linearise these equations, they can be written as a single equation for one of the dynamic variables {(see for example \citet{Sutherland10}), here done for the vertical velocity $w$:}
\begin{equation}\label{eqn:Boussinesq_for_w}
	\left( \pd{^2}{x^2} + \pd{^2}{z^2} \right) \pd{^2 w}{t^2} + N^2 \pd{^2 w}{x^2} = 0,
\end{equation}
where $N$ is the Brunt-Väisälä frequency, $t$ denotes time and $x$ and $z$ are the horizontal and vertical coordinates. As long as $N$ does not depend on $x$ and $t$, the equation admits horizontally and temporally periodic solutions, i.e., 
\begin{equation}
	w(x,z,t) = \hat w (z) \exp(i (kx-\omega t))
\end{equation}
By convention, we consider only $\omega>0$ and focus on $k>0$, i.e., the phase velocity points in the positive $x$-direction. This is no restriction, since the case $k<0$ is completely symmetric to $k>0$. {The partial differential equation \eqref{eqn:Boussinesq_for_w} then can be transformed into the ordinary differential equation}
\begin{equation}\label{eqn:taylor_goldstein_no_wind_global}
	\td{^2\hat w}{z^2} + k^2 \left( \frac{N^2}{\omega^2} - 1\right)\hat w = 0.
\end{equation}
This actually corresponds to a Fourier transformation of equation \eqref{eqn:Boussinesq_for_w} in the horizontal and time coordinates and the resulting equation is widely known as the Taylor-Goldstein equation, which cannot be solved explicitly for non-constant Brunt-Väisälä frequency $N$. But since we are only interested in a confined region within which $N$ is varying, we can approximate $N$ in this region by a piecewise constant function. To be more precise, we are given a function 
\begin{equation}
	N(z) = 
	\begin{cases}
		N_b, & z<z_b \\
		N_{c}(z), & z_b \le z \le z_t \\
		N_t, & z>z_t,
	\end{cases}		
\end{equation}
where $z_b, z_t$ are the bottom and top of the tropopause, respectively (or any region of interest in general), $N_b, N_t$ are constant values of $N$ in the bottom and top layer, respectively, and $N_{c}$ is a (continuous) function of $z$ with $N_{c}(z_b) = N_b$ and $  N_{c}(z_t) = N_t$. Next we partition the continuous part into a piecewise-constant function. This allows us to find explicit solutions in each layer. This method is neither new nor ground-breaking, but the way we use it to compute wave transmission strikes a new path. Moreover, as we will see later, we will gain structural insights into the solution and we are able to give numerical evidence that justifies the ansatz, despite its simplicity.

Let $J$ be a positive integer. For now, $J$ is fixed. Define an equidistant grid of $J$ points from $z_b$ to $z_t$:
\begin{equation}
	z_j = z_b + \frac{j-1}{J-1}(z_t - z_b) \text{ for } j=1, \ldots, J
\end{equation}
and set $N_1 \wdef N(z_1), N_{J+1} \wdef N(z_J)$ and
\begin{equation}
	N_{j} \wdef \frac{N(z_j)+N(z_{j-1})}{2} \text{ for } j = 2,...,J .
\end{equation} 
This can be understood as a piecewise function
\begin{equation}
	\tilde N(z) = N_j, z \in I_j,
\end{equation} 
where $I_j = \left[z_{j-1},z_j\right)$ for $j=2, \ldots,J$, $I_1$ the troposphere region $z<z_b$ and $I_{J+1}$ the stratosphere region $z \ge z_t$. In each single level, we are able to state the Taylor-Goldstein equation, but now $N$ takes a constant value. In particular, for the level $I_j$, we have the equation
\begin{equation}\label{eqn:taylor_goldstein_no_wind_local}
	\td{^2 w_j}{z^2} + k^2 \left( \frac{N_j^2}{\omega^2} - 1\right) w_j = 0.
\end{equation}
Each layer admits explicit plane wave solutions of the form
\begin{equation}\label{eqn:taylor_goldstein_local_solution}
	w_j(z) =   A_j \exp(i m_j z) + B_j \exp(-i m_j z) ,
\end{equation}
where 
\begin{equation}\label{eqn:vertical_wave_number}
m_j = -k \sqrt{\frac{N_j^2}{\omega^2}-1}
\end{equation}
is the vertical wave number and $A_j, B_j$ are the amplitudes of the upward and downward propagating wave, respectively.
Equation \eqref{eqn:vertical_wave_number} is basically the transformed Boussinesq internal gravity wave dispersion relation.
{The representation we use for the wave in equation \eqref{eqn:taylor_goldstein_local_solution} corresponds to the Hilbert transform method for internal gravity wave analysis by \citet{MercierETAL08}. 
Although they were the first to apply this technique in the analysis of observational data, it goes all the way back to the classical works of \citet{EliassenPalm61} and \citet{BookerBretherton67}, who use this representation of plane waves in their theoretical setups. From a mathematical point of view, this demodulation is a rather natural approach to transform a real valued signal into a complex-valued one.
The original real signal then corresponds to the real part of the complex one, in this case the right-hand side of equation \eqref{eqn:taylor_goldstein_local_solution}. This representation has the advantage of tracking upward- and downward-propagating waves. In \citet{MercierETAL08}, however, it is necessary that the stratification is uniform or at most slowly changing. With the multi-layer method we are going to introduce, we will be able to keep track of upward- and downward-propagating waves also in non-uniform background that can change strongly over a fraction of the wavelength. To the best of our knowledge, this is the first technique that is capable of providing this information by construction. 
In the upcoming computation steps, we will work with the Hilbert representation. To obtain the physical solution, we can always take the real part of what we computed. It is important to note that the wave amplitude of the real solution is not computed by taking the real part of the complex amplitude, but also using the imaginary part. That said, a non-vanishing imaginary part of the amplitude basically acts like a phase shift of a cosine. }

Going back to the multi-layer approach, the solution indexed by $1$ references the solution below the tropopause while the solution indexed by $J+1$ corresponds to the solution above it. In particular, the amplitude $A_1$ belongs to the incident wave, while $A_{J+1}$ is the amplitude of the transmitted wave and $B_1$ the one of the reflected wave. Since we are assuming a uniform stratification above the tropopause, there shall be no reflection from upper layers. Hence there is no wave hitting the tropopause from above, i.e., $B_{J+1}=0$. This is a radiation condition which we are going to use later. 

We need to clarify the way we use the indices on the variables that depend on the levels $I_j$. All those variables depend implicitly on the (fixed) number $J$, i.e. for $J_1 \neq J_2$, for example $N_j^{(J_1)} \neq N_j^{(J_2)}$, where the superscript now reflects the dependence on the number of levels. To be precise and keep the variables comparable, one could index them by $\frac{j}{J}$ or superscripting them with the number of steps. But since this is not only cumbersome in writing and reading but also does not provide further benefit (most of the time, we are interested in the variables indexed with $1$ and $J+1$), we omit this dependence but keep it in mind. 

To obtain a solution over the whole domain, we have to match the local solutions at the interfaces in a proper way. Physically meaningful conditions require that the vertical wind speed and the pressure are continuous across the interface (see also \citet{DrazinReid81}). By using the polarisation relations in a Boussinesq fluid (see, e.g., \citet{AKS10} for details), the conditions are equivalent to the requirement that $w_j$ and $ w^\prime_j = \td{w_j}{z}$ are continuous at the interfaces, i.e.,
\begin{subequations}\label{eqnsys:matching_conditions}
\begin{align}
	 w_j(z_j) = w_{j+1}(z_j) \\
	 w_j^\prime (z_j) =  w_{j+1}^\prime (z_j) 
\end{align}	
\end{subequations}
A single pair of the form \eqref{eqnsys:matching_conditions} gives us two equations for the four unknowns $A_j,B_j,A_{j+1},B_{j+1}$. Hence we are able to derive a recurrence relation 
\begin{equation}\label{eqn:recurrence_j_j+1}
\begin{pmatrix}
A_{j+1}\\
B_{j+1}
\end{pmatrix} 
= \mathbold{M}_j
\begin{pmatrix}
A_{j}\\
B_{j}
\end{pmatrix},
\end{equation}
where the matrix $\mathbold{M}_j$ is of the form 
\begin{equation}
\mathbold{M}_j = \begin{pmatrix}
c_j & d_j \\ d_j^* & c_j^*
\end{pmatrix}.
\end{equation}
The particular matrix entries are given by
\begin{subequations}\label{eqnsys:matrix_entires}
\begin{align}
c_j&= \frac{1}{2} \left( \frac{m_{j}}{m_{j+1}}+1 \right) \exp(i(m_j-m_{j+1})z_j)\\
d_j&= -\frac{1}{2} \left( \frac{m_{j}}{m_{j+1}}-1 \right) \exp(-i(m_j+m_{j+1})z_j). 
\end{align}
\end{subequations}
For later reference, we introduce the $^*$-operation which changes the sign of the argument of the $\exp$-function, i.e., 
\begin{subequations}\label{eqnsys:matrix_entires_star}
\begin{align}
c_j^*&= \frac{1}{2} \left( \frac{m_{j}}{m_{j+1}}+1 \right) \exp(-i(m_j-m_{j+1})z_j)\\
d_j^*&= -\frac{1}{2} \left( \frac{m_{j}}{m_{j+1}}-1 \right) \exp(i(m_j+m_{j+1})z_j). 
\end{align}
\end{subequations}
As long as $m_j$ and $m_{j+1}$ are real-valued, this corresponds to complex conjugation. Imaginary values for $m$ occur only when the waves are encountering a region of decreased stratification, where $N < \omega$. We will see later that these cases are harder to deal with when trying to find a limit for an increasing number of layers, hence they have to be treated very carefully.

%

We can state a relation like equation \eqref{eqn:recurrence_j_j+1} for all $j = 1, \ldots, J$ and combine them to obtain a chain of equations:
\begin{equation}\label{eqn:matrix_prod}
\begin{pmatrix}
A_{J+1}\\
B_{J+1}
\end{pmatrix} 
= \mathbold{M}_J
\begin{pmatrix}
A_{J}\\
B_{J}
\end{pmatrix} = \mathbold M_J \mathbold{M}_{J-1}
\begin{pmatrix}
A_{J-1}\\
B_{J-1}
\end{pmatrix} = \ldots =
\underbrace{\prod_{k=J}^1 \mathbold M_k}_{\wdefi \mathbold{M}} 
\begin{pmatrix}
A_{1}\\
B_{1}
\end{pmatrix}
\end{equation}
We have to be careful about the order of the matrix multiplication, since it is in general not commutative.

Now we need to recall what the different amplitudes with their respective indices represent.
$A_{J+1}$ is the transmitted wave, $A_1$ is the incident wave and $B_{J+1} = 0$, i.e., there is no downward propagating wave in the uppermost layer. To compute a transmission coefficient, we have to relate $A_1$ and $A_{J+1}$. We have 
\begin{align}
A_{J+1} &= M_{1,1} A_1 + M_{1,2} B_1 \\
0 &= M_{2,1} A_1 + M_{2,2} B_1,
\end{align}
where $M_{k,l}$ are the entries of $\mathbold M$. Solving the equation system for $A_1$ and $A_{J+1}$ shows that 
\begin{equation}\label{eqn:amplitude_relation}
\frac{A_{J+1}}{A_1} = \left( M_{1,1} - \frac{M_{1,2}M_{2,1}}{M_{2,2}} \right) = \frac{\operatorname{det}(\mathbold M)}{M_{2,2}}.
\end{equation}
To compute a meaningful transmission coefficient, we need to find a quantity that is conserved over the whole domain. Since we did not allow for dissipation or background horizontal wind in the equations, wave energy (sometimes called perturbation energy) is conserved. It consists of kinetic and potential energy. Moreover, we are in a horizontally periodic domain, so it is convenient to have a look at the horizontally averaged energy
\begin{equation}
\langle E \rangle = \langle E_{kin} \rangle + \langle E_{pot} \rangle = \frac{1}{2} \rho_b \frac{N^2}{\omega^2} \labs A_w \rabs^2.
\end{equation}
Here $\rho_b$ is the background density and $A_w$ is the amplitude of the vertical wind. The unit of $\langle E \rangle$ is energy per unit volume, therefore the correct term would be energy density. Energy can be derived from this expression by integrating over a fixed control volume. But since there is no danger of confusion, we stick to the term "energy" for $\langle E \rangle$. 
The conservation equations for kinetic and potential energy can be derived directly from the Boussinesq equations. Horizontal averaging and adding the equations yield
\begin{equation}
 \pd{\langle E \rangle}{t} + \pd{\langle \mathcal{F}_z \rangle}{z} = 0,
\end{equation}
where $\langle \mathcal{F}_z \rangle = c_{g_z} \langle E \rangle$ is the vertical wave energy flux and $c_{g_z} = \pd{\omega}{m}$ denotes the vertical component of the group velocity. A full derivation of the above equations can be found in chapter $3.4$ of \citet{Sutherland10}.

The average energy at a fixed location does not change in time, since we assumed the solution to be periodic in time. The remaining term, namely 
\begin{equation}\label{eqn:vert_flux_const}
\pd{\langle \mathcal{F}_z \rangle}{z} = 0
\end{equation}
basically says that the vertical mean wave energy flux is constant. Again, we use the work of \citet{Sutherland10} to find out that this flux for a Boussinesq wave in uniform stratification is given by 
\begin{equation}
\langle \mathcal{F}_{z} \rangle = \frac{1}{2} \rho_b \frac{N^3}{\omega^2 k} \sin(\alpha) \cos^2(\alpha) \labs A_w \rabs^2
\end{equation} 
where $\alpha = \arctan(\frac{m}{k})$ is the angle between the wave vector and the horizontal. Using the identities 
\begin{align}
	\sin(\arctan(x)) &= \frac{x}{\sqrt{1+x^2}}\\
	\cos(\arctan(x)) &= \frac{1}{\sqrt{1+x^2}}
\end{align}
and the internal gravity wave dispersion relation
\begin{equation}
\omega = \frac{Nk}{\sqrt{k^2+m^2}}
\end{equation}
we obtain that
\begin{equation}
\langle \mathcal{F}_{z} \rangle =\frac{\rho_b m \omega}{2k^2} \labs A_w \rabs^2.
\end{equation}
Since we have a region with uniform stratification below and above the tropopause, we can compare the upward energy fluxes in both of those regions. Recall that $\omega$ and $k$ are chosen to be constant. Moreover, we made the assumption that the density does not vary too much over the tropopause, so that we take a reference value $\varrho_0$ for both regions. The transmission coefficient is then defined as the ratio of the upward energy flux above and below the tropopause
\begin{equation}\label{eqn:TC}
TC \wdef \frac{\langle \mathcal{F}_{z} \rangle_\text{above,up}}{\langle \mathcal{F}_{z} \rangle_\text{below,up}}=\frac{m_{J+1}}{m_1} \labs\frac{A_{J+1}}{A_1}\rabs^2 = \frac{m_{J+1}}{m_1} \labs \frac{\det(M)}{M(2,2)} \rabs^2.
\end{equation}
In a similar fashion we can define a reflection coefficient, which compares the upward flux with the downward flux below the tropopause:
\begin{equation}
RC \wdef \labs \frac{B_1}{A_1}\rabs^2
\end{equation}
By conservation of vertical energy flux, given by equation \eqref{eqn:vert_flux_const}, we have that 
\begin{equation}
TC+RC =1.
\end{equation}


\section{Limit behaviour}
\label{sec:convergence}

{
This section is devoted to the investigation of an increasing number of layers, eventually tending to infinity. First, we have a view on how the transmission coefficients for a given parameter set change when the number of layers is altered. Then, we will be able to derive an expression for the number of layers tending to infinity, which results in a reformulation of equation \eqref{eqn:taylor_goldstein_no_wind_global} in a set of variables that allows the distinction between up- and downward propagating wave modes. This immediately gives rise to transmission and reflection coefficients for an infinite number of layers. We also investigate how quick the multi-layer method converges to this limit. 

It should be mentioned that the multi-layer method itself can be used to construct an approximate solution to equation \eqref{eqn:taylor_goldstein_no_wind_global}. When reformulating equation \eqref{eqn:taylor_goldstein_no_wind_global} to a first-order system, where one of the variables corresponds to the first derivative of the solution, the multi-layer method with the corresponding matching conditions \eqref{eqnsys:matching_conditions} can be written as a one-step method, similar what is done in \citet{Lara04}, where also a proof of convergence is given. This puts the multi-layer method on mathematically solid ground.}

\subsection{Behaviour with increasing discretisation levels}

\begin{figure}
\includegraphics[width=0.497\textwidth,keepaspectratio]{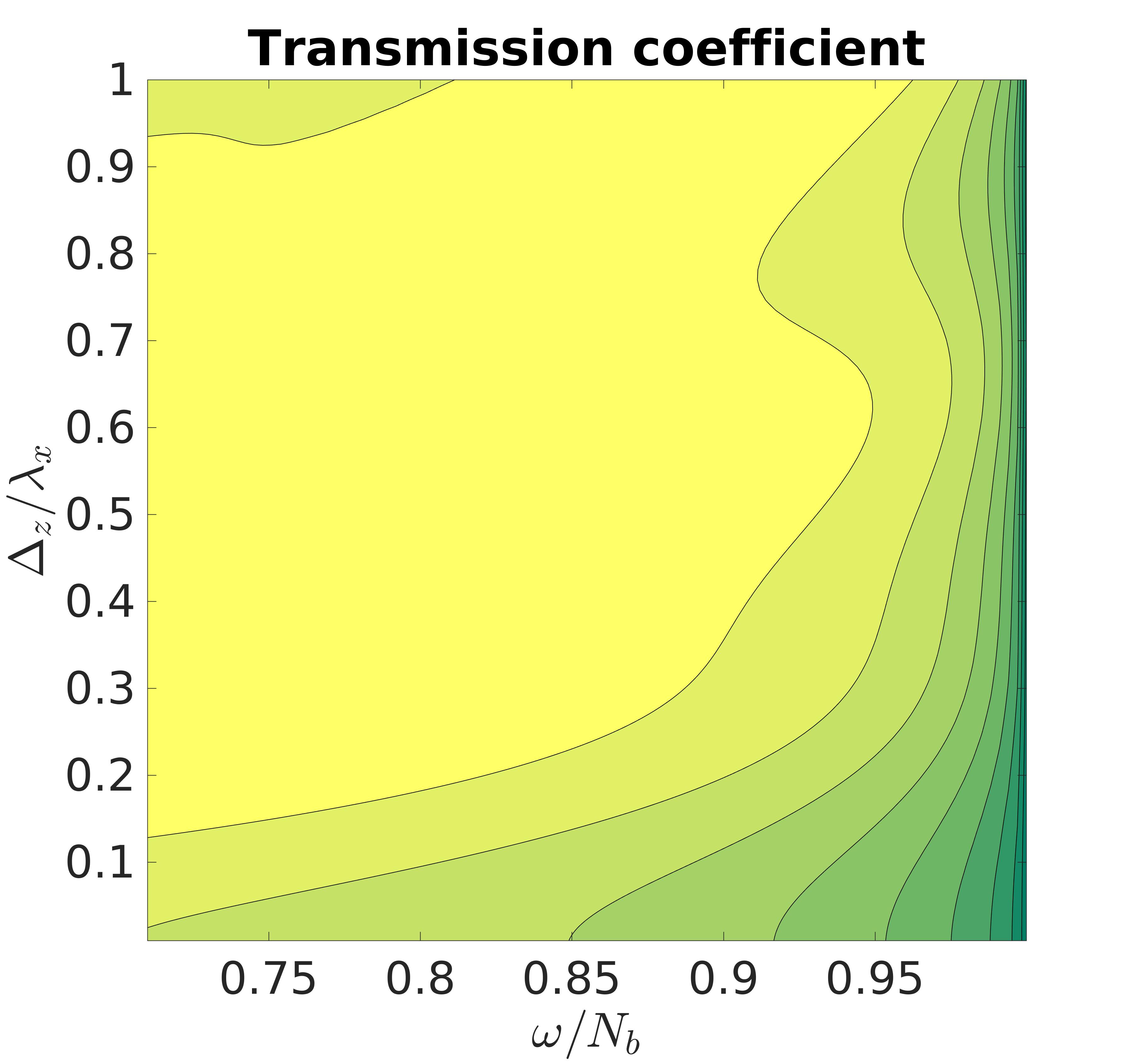}
\includegraphics[width=0.497\textwidth,keepaspectratio]{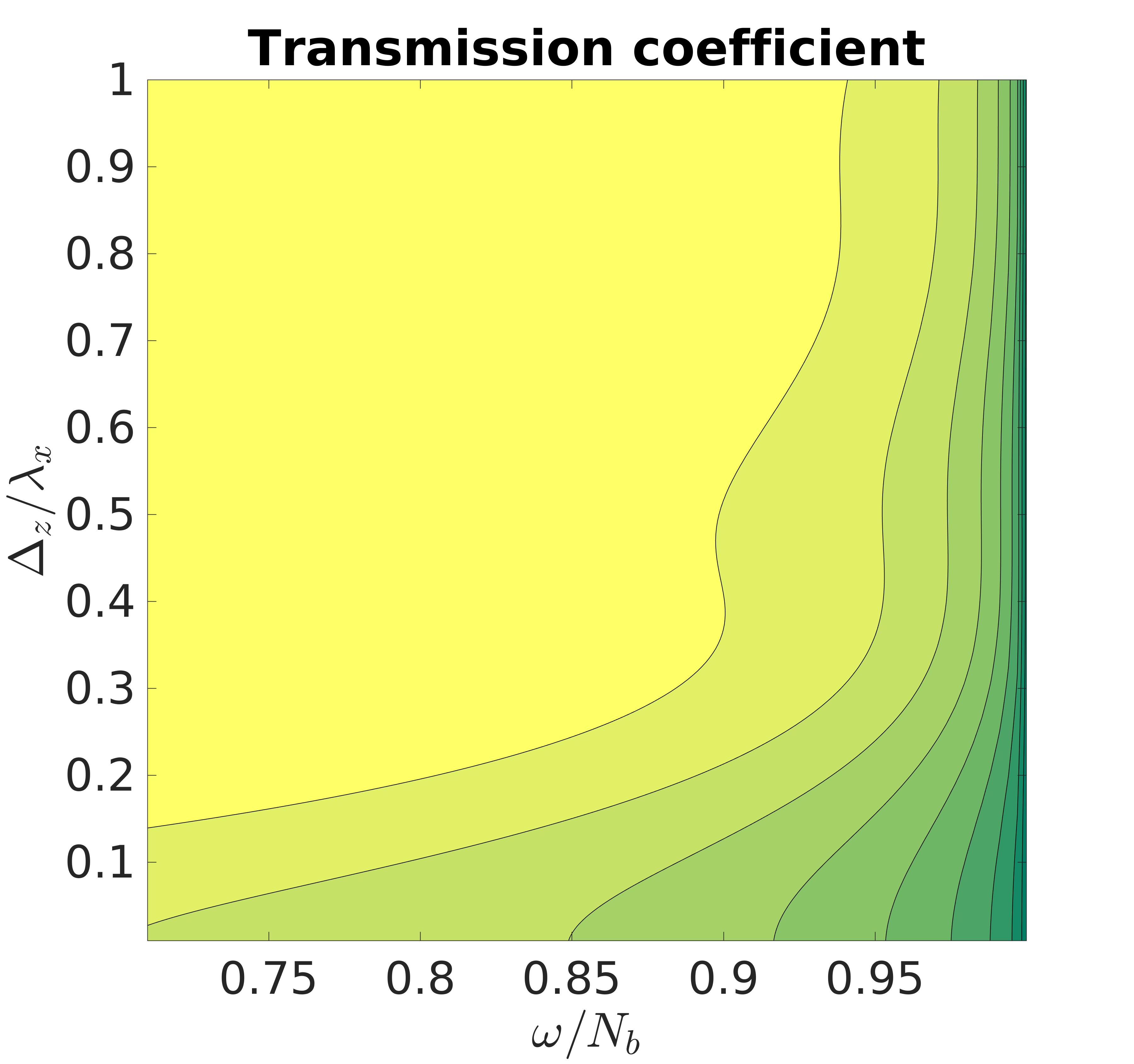}
\includegraphics[width=0.497\textwidth,keepaspectratio]{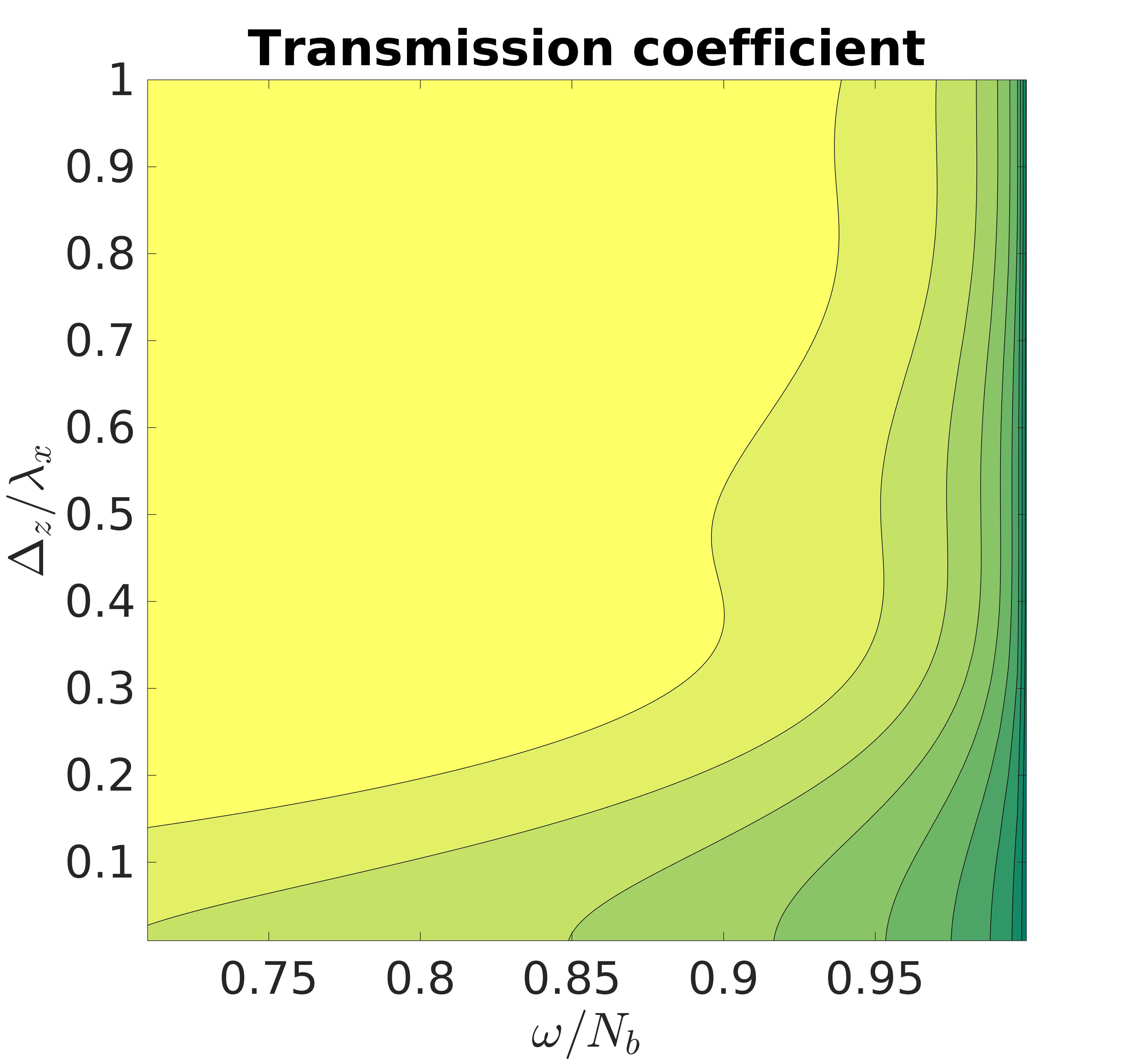}
\includegraphics[width=0.497\textwidth,keepaspectratio]{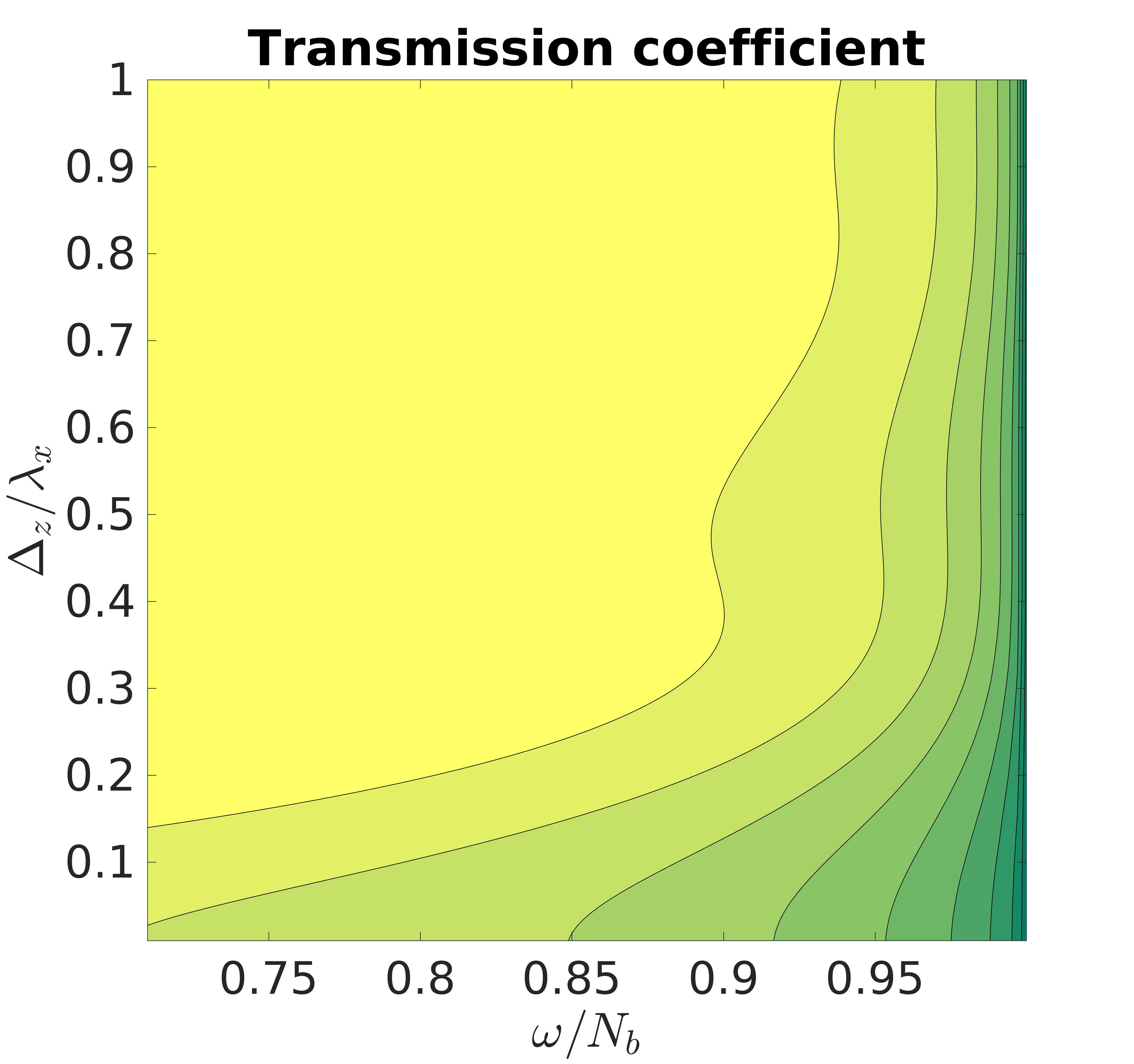}
\includegraphics[width=\textwidth, keepaspectratio]{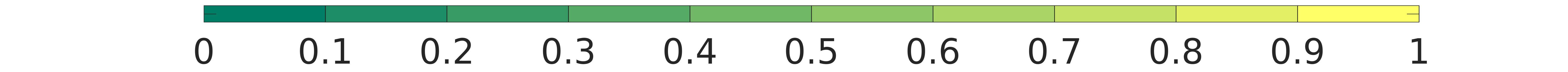}
\caption{The panels show the transmission coefficient for profile \eqref{profile:linear_increasing} for given horizontal wavelength $\lambda_x$ and frequency $\omega$ for different number of layers $J$: { $J=4$ on top-left, $J=16$ on top-right, $J=64$ on bottom-left and $J=256$ on bottom-right.} {Except for $J=4$}, the pictures are nearly indistinguishable. As we will see in figure \ref{fig:error}, we have a convergence rate of about 2.} 
\label{fig:increasing_j}
\end{figure}
Our first step towards showing that the method is convergent with increasing number of layers is to have a look at how the transmission coefficient is influenced by this very number, again denoted by $J$. As a reference setup, we take a stratification profile that increases linearly from a value $N_b$ to a value $N_t>N_b$ over a confined region $[ z_b,z_t]$ with length $\Delta_z \wdef z_b-z_t$:
\begin{equation}\label{profile:linear_increasing}
	N(z) = 
	\begin{cases}
	N_b, & z<z_b \\
	N_b + \frac{z-z_b}{z_t-z_b}(N_t-N_b), & z_b \le z \le z_t\\
	N_t,& z_t < z.
	\end{cases}
\end{equation}
A visualisation of the profile can be seen in the right panel of figure \ref{fig:linear_profile_results}. Our computation domain in this investigation covers a frequency range from $0$ to nearly $N_b$ and a wavelength range from $100 \Delta_z$ to about $ \Delta_z$.  The results for various values of $J$ can be seen in figure \ref{fig:increasing_j}. The images show a colour plot of the transmission coefficient, where the x-axis corresponds to frequency and y-axis to horizontal wavelength. Here, we are not interested in the interpretation of the images in themselves, but rather in the transition from small to large $J$. As one can see, except for the case $J=4$ (upper left panel), one can barely spot any difference in the pictures. This leads to the hypothesis that the method converges to a certain limit for increasing $J$, and from what we see qualitatively, this convergence seems to be pretty fast. The next step will be to find the limit.

\subsection{Limit solution}
Unfortunately, finding the limit has turned out to be much harder than anticipated. Since the matrices $\mathbold{M}_j$ have complex entries, {there seems to be no closed formula for} 
\begin{equation}
\lim_{J \to \infty} \prod_{k=J}^1 \mathbold M^{(J)}_k,
\end{equation}
because most {known formulas blow up because of} the non-vanishing imaginary part (under certain conditions, we managed to find a formula, but in general, these conditions cannot be fulfilled). Hence, a new approach was needed.

{The idea is to reformulate the limit process as a differential equation for a vector consisting of the amplitudes for the upward and downward propagating wave.} We know that the depth of each layer is $h=\frac{\Delta_z}{J}$, so the limit process $J \to \infty$ can also be seen as $\frac{\Delta_z}{J} \to 0$ or $ h \to 0$. Moreover, the $j$-indexed variables $A_j,B_j,m_j$ are approximations of their continuous counterparts at $z_j$. In fact, they are approximations at $z_j + \frac{h}{2}$, which goes to $z_j$ in the limit $h \to 0$. Also, a reformulation such that e.g. $m_j = m(z_j)$ gives the same result. By using the recurrence relation \eqref{eqn:recurrence_j_j+1} we can write
\begin{equation}
\begin{pmatrix}
A_{j+1}\\
B_{j+1}
\end{pmatrix} -
\begin{pmatrix}
A_j\\
B_j
\end{pmatrix}=\left(  \mathbold M^{(J)}_j-I \right)
\begin{pmatrix}
A_{j}\\
B_{j}
\end{pmatrix}  , 
\end{equation}
where $I$ is the 2-by-2-identity matrix. Dividing now by $h$ and taking the limit $h \to 0$, the left-hand side converges to the $z$-derivative of the amplitudes. Using the short-hand notation $\mathcal{A} {= \mathcal{A}(z)} =  \begin{pmatrix}
A(z)\\
B(z)
\end{pmatrix} $ for the vector of amplitudes, we have
\begin{equation}
\td{ \mathcal{A} }{z} =  \lim_{h \to 0} \frac{ \left(  \mathbold M^{(J)}_j - I \right)}{h} \mathcal{A}.
\end{equation}
This is now a differential equation for the amplitudes in $\mathcal{A}$. If we want to have any chance of solving it (either analytically or numerically), we have to execute the limit process 
\begin{equation}\label{eqn:limit}
\lim_{h \to 0} \frac{ \left(  \mathbold M^{(J)}_j - I \right)}{h}.
\end{equation}
This is done component-wise. The upper-left entry of the matrix inside the limit in \eqref{eqn:limit} is $\frac{c_j-1}{h}$. If we let $h \to 0$, we get 
\begin{equation}\label{eqn:limit_matrix_diagonal_entry}
f(z) \wdef \lim_{h \to 0} \frac{c_j-1}{h} = -\frac{m^\prime(z)}{2m(z)} -im^\prime(z)z.
\end{equation}
With a similar computation for the off-diagonal entry $\frac{d_j}{h}$, we obtain 
\begin{equation}\label{eqn:limit_matrix_off_diagonal_entry}
g(z) \wdef \lim_{h \to 0} \frac{d_j}{h} = \frac{m^\prime(z)}{2m(z)} \exp(-2im(z)z).
\end{equation}
The full derivation for the limit in equations \eqref{eqn:limit_matrix_diagonal_entry} and \eqref{eqn:limit_matrix_off_diagonal_entry} can be found in the appendix. 
The respective limits for the starred entries yield the same except for a replacement of $i$ by $-i$. Hence the differential equation for the amplitudes is
\begin{equation}\label{eqn:limit_ode_matrix_form}
\td{ \mathcal{A} }{z} =  \underbrace{\begin{pmatrix}
f(z) & g(z) \\ g^*(z) & f^*(z)
\end{pmatrix}}_{\wdefi \mathcal{M}(z)} \mathcal{A}.
\end{equation}
This equation can only be solved analytically if $\mathcal{M}(z_1)\mathcal{M}(z_2)=\mathcal{M}(z_2)\mathcal{M}(z_1)$ holds for all $z_1,z_2$ in the integration domain. Unfortunately, this is in general not true for arbitrary stratification profiles. 
{What we do know, however, is that if the functions are analytic over the interval $[z_b,z_t]$, equation \eqref{eqn:limit_ode_matrix_form} has a unique analytic solution for arbitrary initial data $\mathcal{A}(z_0) = \mathcal{A}_0$, $z_0 \in [z_b,z_t]$. See, for example, \citet{Teschl12} for the theory on complex ODEs. It is easy to check that this is the case for stratification profiles and wave parameters such that there is no reflection layer, i.e. a point $z_r$ where $N(z_r) = \omega$. Although securing the existence of solutions, finding analytic or even explicit expressions for them will be a nearly hopeless undertaking. Another approach to {finding} at least approximate solutions are power series methods. Since analytic functions on an open subset coincide locally with a convergent power series (see , e.g., \citet{Stalker98} for details), we can make a power series ansatz for the solution of equation \eqref{eqn:limit_ode_matrix_form}. In order to do so, we extend $[z_b,z_t]$ to an open subset of the complex numbers, in which $\mathcal{M}$ is still analytic. But although the matrix has no singularities for real values, it has some for certain complex numbers, which drastically restricts the radius of convergence of the power series solution to a value that is not guaranteed to be large enough to cover the whole region of interest. However, it would be possible to partition the interval into smaller segments and finding the power series solution in each segment, but this procedure is very tedious and still only yields a solution up to a certain precision. We will see in the upcoming error analyses that the multi-layer method yields very accurate results notwithstanding that it is a much easier-to-apply technique. Hence, the evaluation of equation \eqref{eqn:limit_ode_matrix_form} will be done numerically.

This leaves us with another challenge. If the stratification profile contains a reflection layer $z_r$ such that $N(z_r)=\omega$, the entries of the coefficient matrix $\mathcal{M}$ tend to infinity, because $m(z_r)=0$. By regarding equation \eqref{eqn:limit_ode_matrix_form} as a system of complex differential equations, the point $z_r$ is an isolated singularity. At first sight, it seemed to us that the singularity is a first order pole, since functions of the form $\frac{f^\prime}{f}$, where $f$ has a zero of any order at some point $z_0$ do have a first order pole at $z_0$. Unfortunately, the extra terms that are prevalent in the coefficients, i.e. $im^\prime(z)z$ and $\exp(-2im(z)z)$ involve square roots, which results in $f$ and $g$ not being holomorphic in a punctured disk around $z_r$, since the complex square root has two branches and therefore it is only holomorphic in a disk with one half-axis removed. Therefore, all theorems about existence and  structure of solutions can not be applied. Existing research in this case, such as \citet{SutherlandYewchuck04}, who investigate propagation of gravity waves through a layer of sudden reduced or vanishing stratification, suggest some sort of "wave tunnelling" (a term coined by the comparison to quantum tunnelling of electrons in quantum physics) through this region, dependent on the wavelength of the incident wave. The findings from our multi-layer method confirm those results also for continuous transitions to a lower value of the Brunt-Väisälä frequency (see section \ref{subsec:tunnelling}). Hence, an intensive investigation of the behaviour of gravity waves near reflection layers would require a scale analysis for different regimes of vertical wavelengths. This task is taken on by the authors, but lies outside the scope of this paper.  }


\subsection{Error analysis}\label{subsec:error_analysis}

This subsection is dedicated to the numerical integration of equation \eqref{eqn:limit_ode_matrix_form}. {To this end}, proper boundary conditions are needed. Since we are interested in waves that are initially travelling upwards and encountering a non-uniform stratification over a confined region and eventually reaching a region of uniform stratification again, we require that there is only a wave incident on the non-uniform region from below. Hence, above this region, there is no wave that is travelling downwards. Moreover, we are only interested in ratios between the incident and the transmitted wave. Therefore we are free to choose the value of the incident or the transmitted wave, {since the equation is linear}. To be more precise, if $z_t$ is the top of the non-uniform region, then we take the boundary conditions
\begin{equation}\label{eqn:ODE_amplitudes}
 \mathcal{A}(z_t) = \begin{pmatrix}
 A(z_t) \\ B(z_t)
 \end{pmatrix} = \begin{pmatrix}
 1 \\ 0
 \end{pmatrix}.
\end{equation}

We want to analyse the error between the limit solution and the multi-layer solution, which will give strong evidence of the correctness of our method. As model setup, we again choose the linearly increasing profile \eqref{profile:linear_increasing}. Our analysis consists of two parts. The first one is an error computation over a large domain of wave numbers and frequencies while keeping the number of discretisation levels constant at $J=512$ to show that the error is small over the whole wave number-frequency-domain. The second one chooses several specific points in this domain and tracks the error for an increasing number of levels $J$, up to $J=10^5$.

\begin{figure}
\includegraphics[width=0.497\textwidth, keepaspectratio]{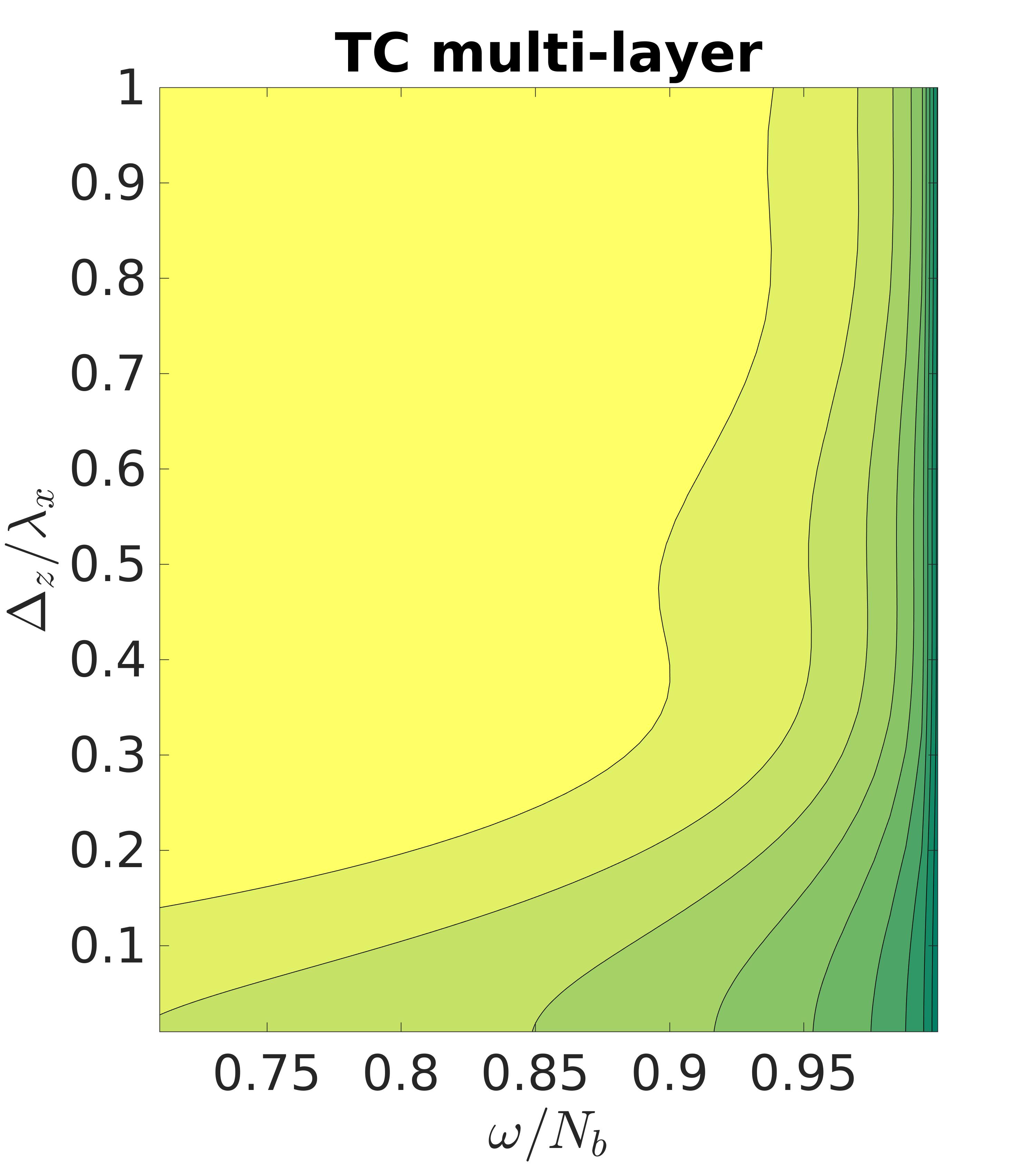}
\includegraphics[width=0.497\textwidth, keepaspectratio]{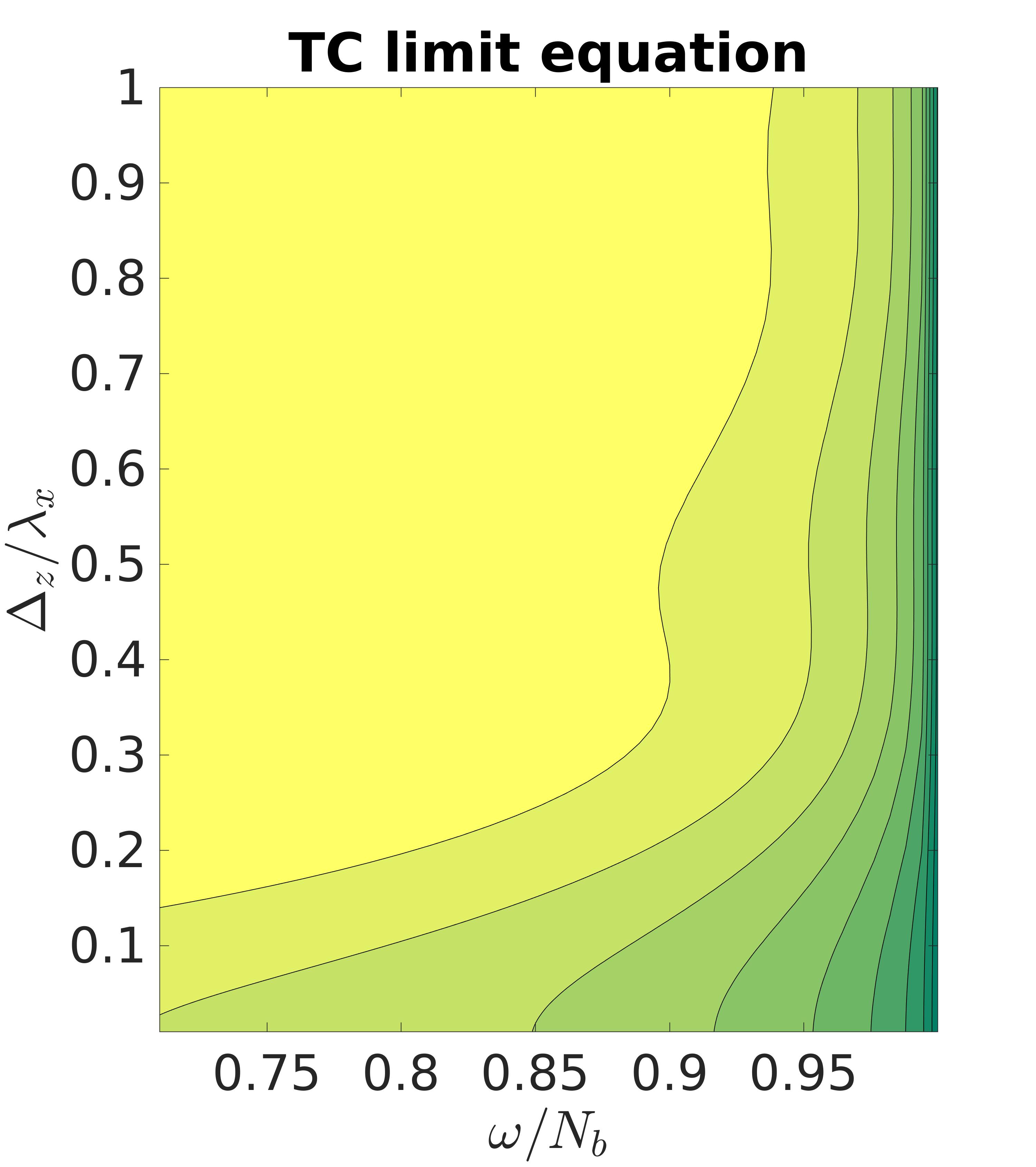}
\includegraphics[width=\textwidth, keepaspectratio]{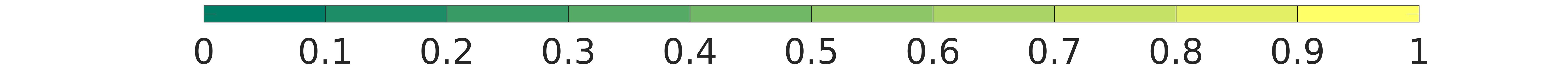}
\caption{The left panel shows the multi-layer method with $J=512$ steps and the right panel shows the values of the transmission coefficient computed from the numerical evaluation of the limit approach. With bare eye, they are indistinguishable. At every single point in the domain, the error is smaller than $10^{-5}$}
\label{fig:comparison_discrete_limit}
\end{figure}

\begin{figure}
		\subfigure[]{\includegraphics[width=0.325\textwidth, keepaspectratio]{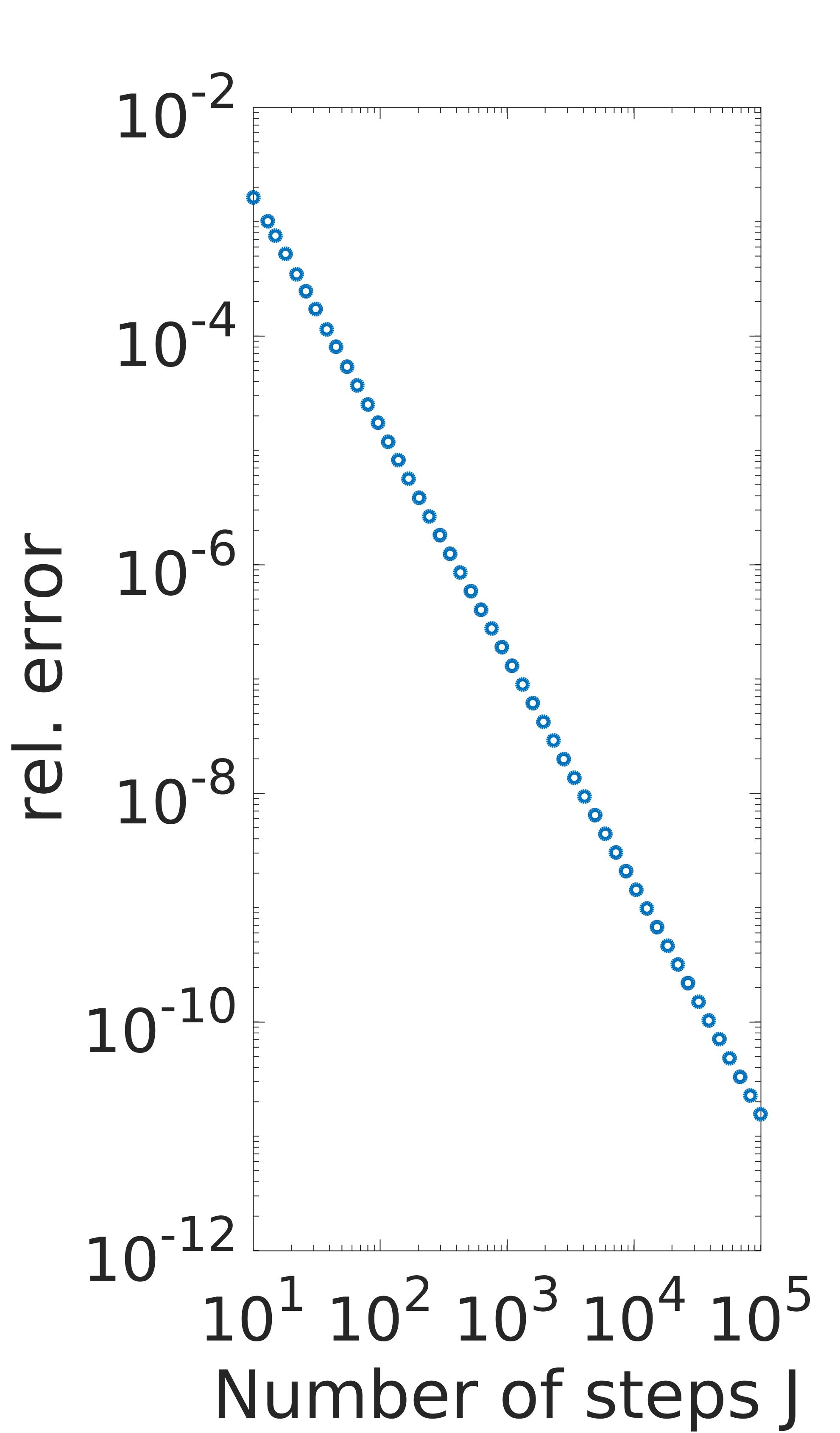}}
		\subfigure[]{\includegraphics[width=0.325\textwidth, keepaspectratio]{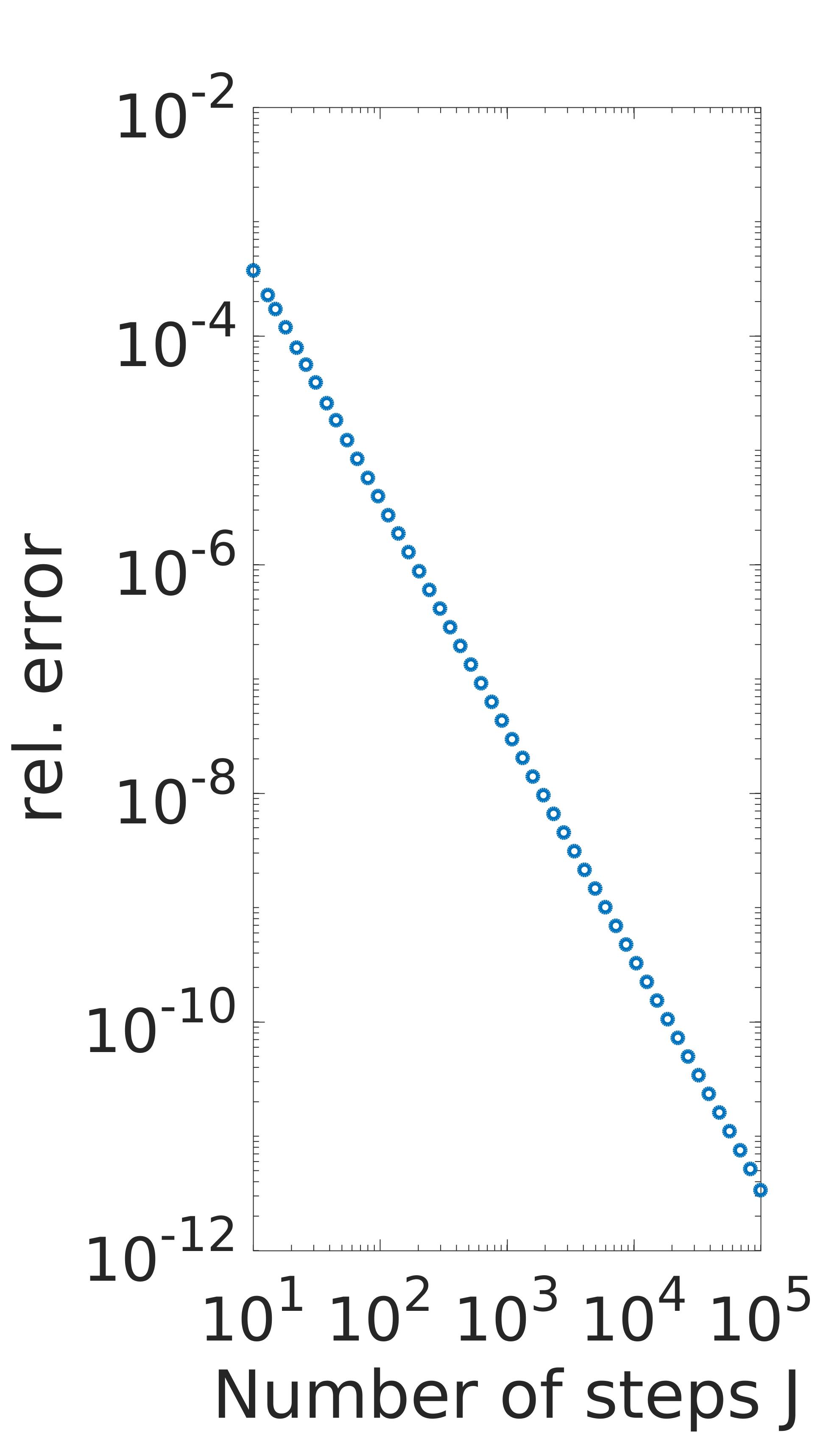}}
		\subfigure[]{\includegraphics[width=0.325\textwidth, keepaspectratio]{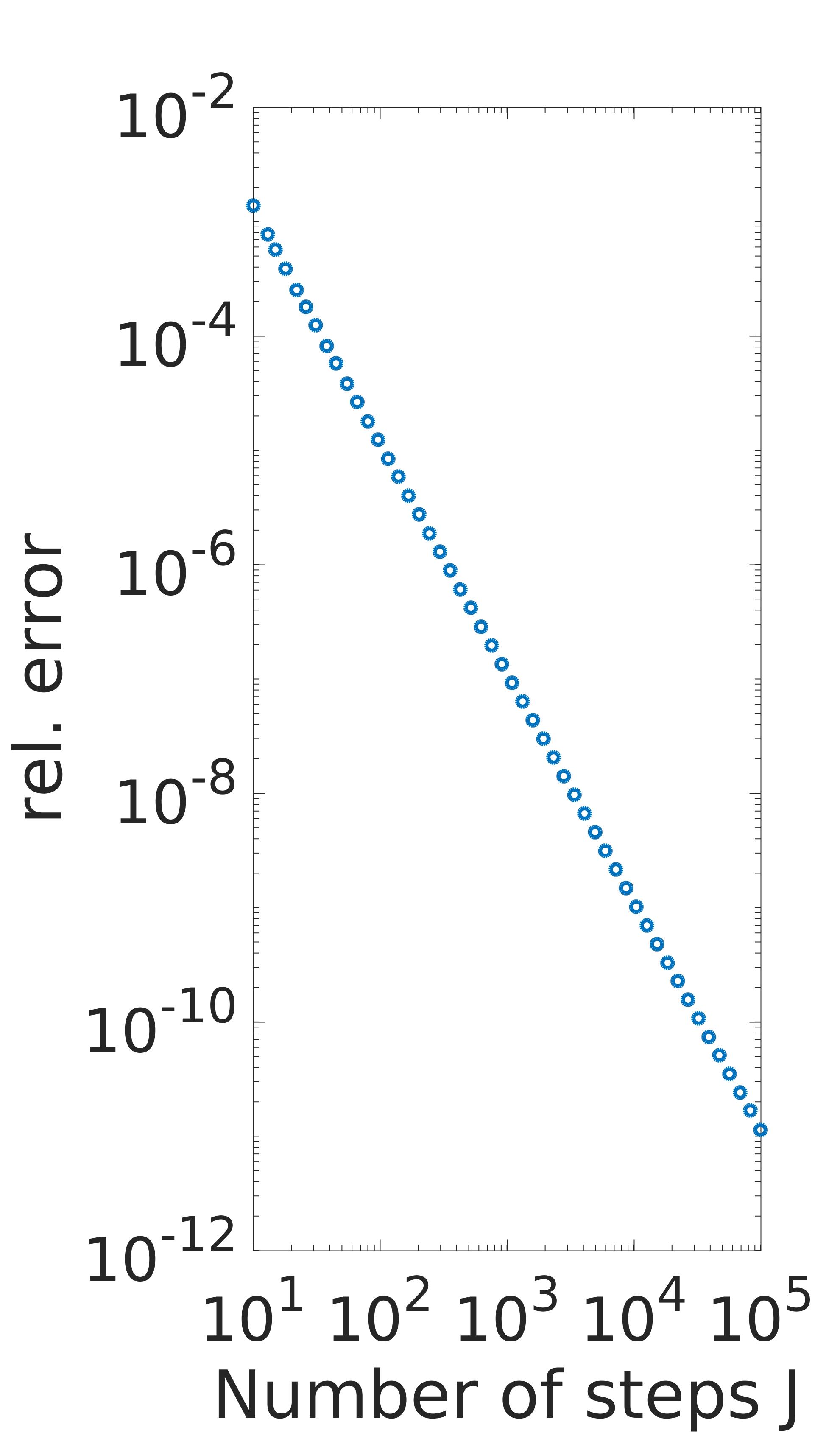}}
	\caption{Relative error for profile \eqref{profile:linear_increasing} with $N_t = 2 N_b$, $\omega_0 = \frac{N_b}{\sqrt{2}}$ and different horizontal wavelengths $\lambda_x$: { $\lambda_x = \Delta_z$ in panel (a), $\lambda_x = 2 \Delta_z$ in (b) and $\lambda_x = 10 \Delta_z$ in (c)}. The mean slope of the different plots is nearly the same, namely $\mu \approx -2$. This means that the relative error decreases quadratically with the number of steps. }
	\label{fig:error}
\end{figure}

The results of the first part can be seen in figure \ref{fig:comparison_discrete_limit}. The pictures show the transmission coefficient for a certain domain of wavelengths and frequencies, similar to figure \ref{fig:increasing_j}. But we are still not interested in the meaning of the pictures on their own but in the comparison of both. For the left sub-figure, we derived the transmission coefficient from the multi-layer method with $J=512$ levels, the right panel shows the transmission coefficient computed from solving equation \eqref{eqn:limit_ode_matrix_form} numerically. It is impossible to spot any difference between the two frames. Computing the relative error yields the estimate
\begin{equation}
	\max_{\omega, \lambda_x} \frac{\labs TC_{d}(\omega,\lambda_x) - TC_l(\omega,\lambda_x)\rabs  }{\labs TC_l(\omega,\lambda_x) \rabs}< 7 \cdot 10^{-6}
\end{equation} 
For the second analysis, we fix specific wave parameters, i.e, a pair $(\lambda_{x,0},\omega_0)$ of wavelength and frequency, and analyse how the relative error develops for increasing $J$. In particular, we perform the calculation for three different wavelength-frequency-pairs. We choose $\omega_0 = \frac{N_b}{\sqrt{2}}$ for all three cases and have a look at the wavelengths $\Delta_z, 2 \Delta_z$ and $10 \Delta_z$. The results can be seen in figure \ref{fig:error}. 

We computed the relative error of the limit solution and the discrete solution for several numbers of layers $J$, that were logarithmically spaced between $10^1$ and $10^5$. For any two adjacent points, we computed the slope in the $\log$-$\log$ diagram and for every wavelength, the mean and the standard deviation of all computed slopes. We found the mean slopes to be $\mu_a = -2.0050 \pm 0.0314, \mu_b = -2.0073 \pm 0.0533$ and $\mu_c = -2.0269 \pm 0.1184$, where the indices correspond to the respective sub-plots of figure \ref{fig:error}. For case c, the last few values are around the tolerance of the numerical scheme, hence there are somewhat larger fluctuations. But nonetheless, we observe that in all three test cases, the error decreases quadratically with the number of steps, until the error reaches the region where the tolerance of the scheme and the machine precision prevent a more precise computation. Even for a coarse discretisation with $J =100$ layers, the relative error is about $10^{-5}$. Based on this, we will use $J=128$ layers for the forthcoming computations. This guarantees fast run times as well as results that are sufficiently accurate.


\section{Results}
\label{sec:results}

In this section, we present some results that are obtained for several stratification profiles. First, we focus on the linearly increasing profile we already used for some model computations. Afterwards, we show that the model also supports wave tunnelling that was already described by \citet{SutherlandYewchuck04}. At the end, we present results for a profile that has characteristics of a real tropopause. All profiles share a common structure, namely that we have a region of non-constant stratification of depth $\Delta_z$ that has a region of constant stratification with value $N_b$ below and with value $N_t$ above it. This is about what we can observe in the atmosphere: The stratification changes rapidly in the tropopause and is nearly constant in the free troposphere and the stratosphere. We will focus on a frequency range from $0$ to $N_b$, since waves with frequencies larger than $N_b$ are evanescent. The wavelength spectrum differs in the various examples but ranges in general from $ \Delta_z$ to $100 \Delta_z$.{  We use a parameter space of 300 horizontal wave numbers and 300 frequencies, resulting in a total of 90000 grid points. The number of layers equals 128. The computations are performed on the first author's office computer with a standard Intel\textsuperscript{\textregistered} Core\texttrademark\ i7-3770 CPU and 8 GB RAM. The software we used is \textsc{Matlab}. The computation times lie within a range of $70$ to $80$ seconds when computed on a single core. Compared to the numerical method of \citet{NaultSutherland07}, who need about 1 day to simulate $300 \times 300$ parameters on a "typical desktop computer" at that time, this is a decrease in computation time by a factor of about 1000. Even when considering the slightly higher clock rate and RAM, the multi-layer method is much more efficient. }

\subsection{Linear increase}

\begin{figure}
\includegraphics[width=0.395\textwidth, keepaspectratio]{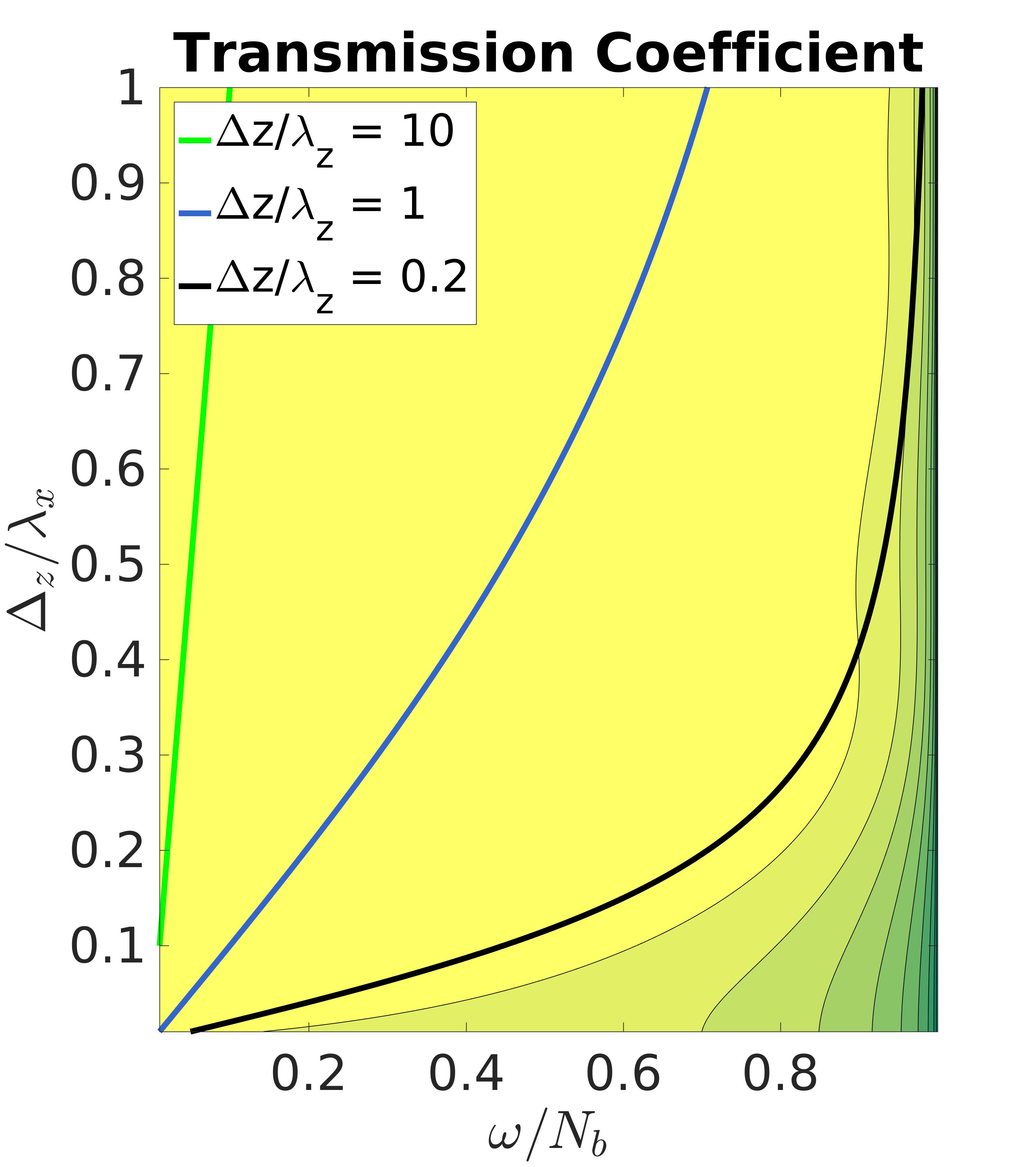}
\includegraphics[width=0.395\textwidth, keepaspectratio]{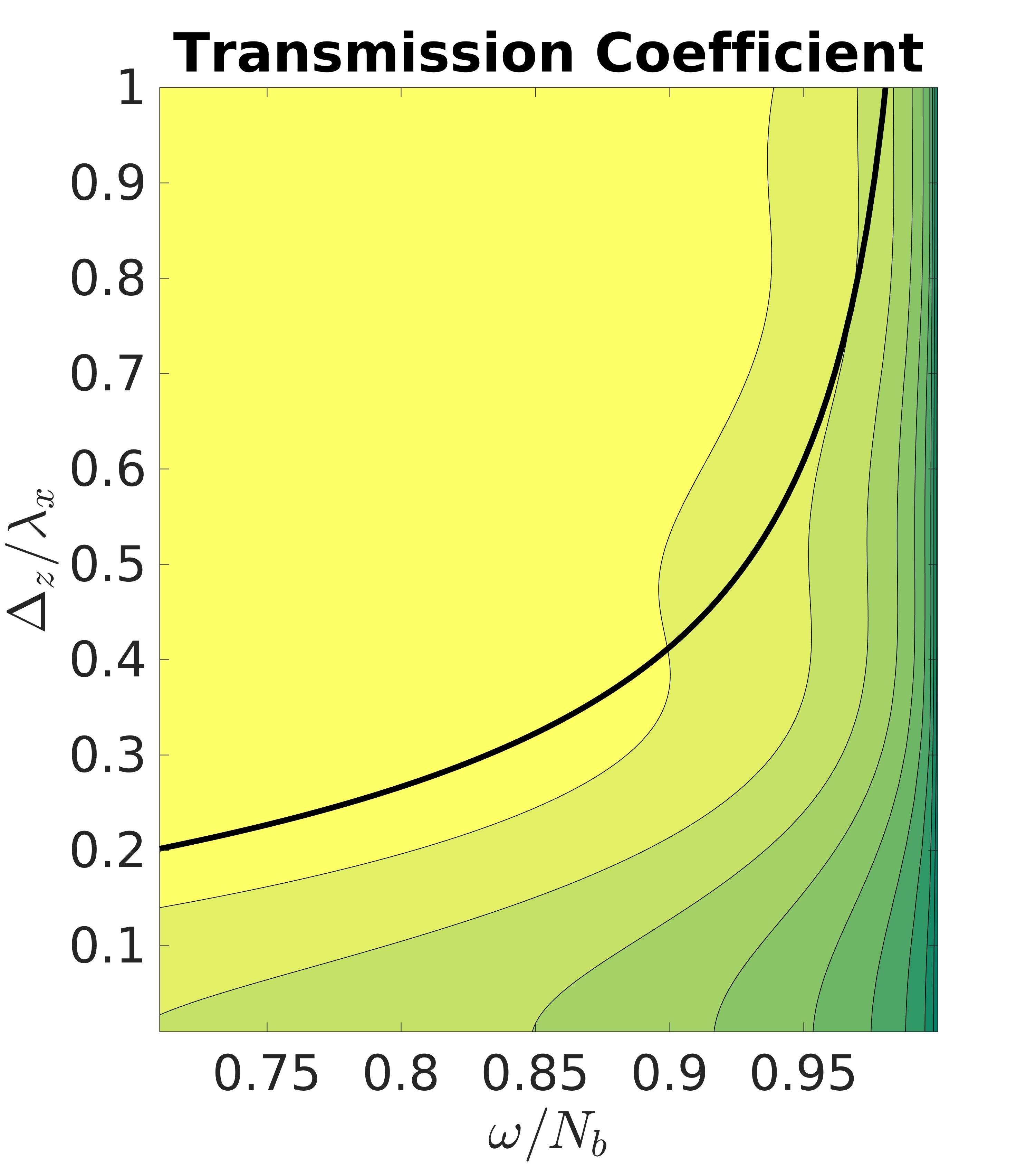}
\includegraphics[width=0.1975\textwidth, keepaspectratio]{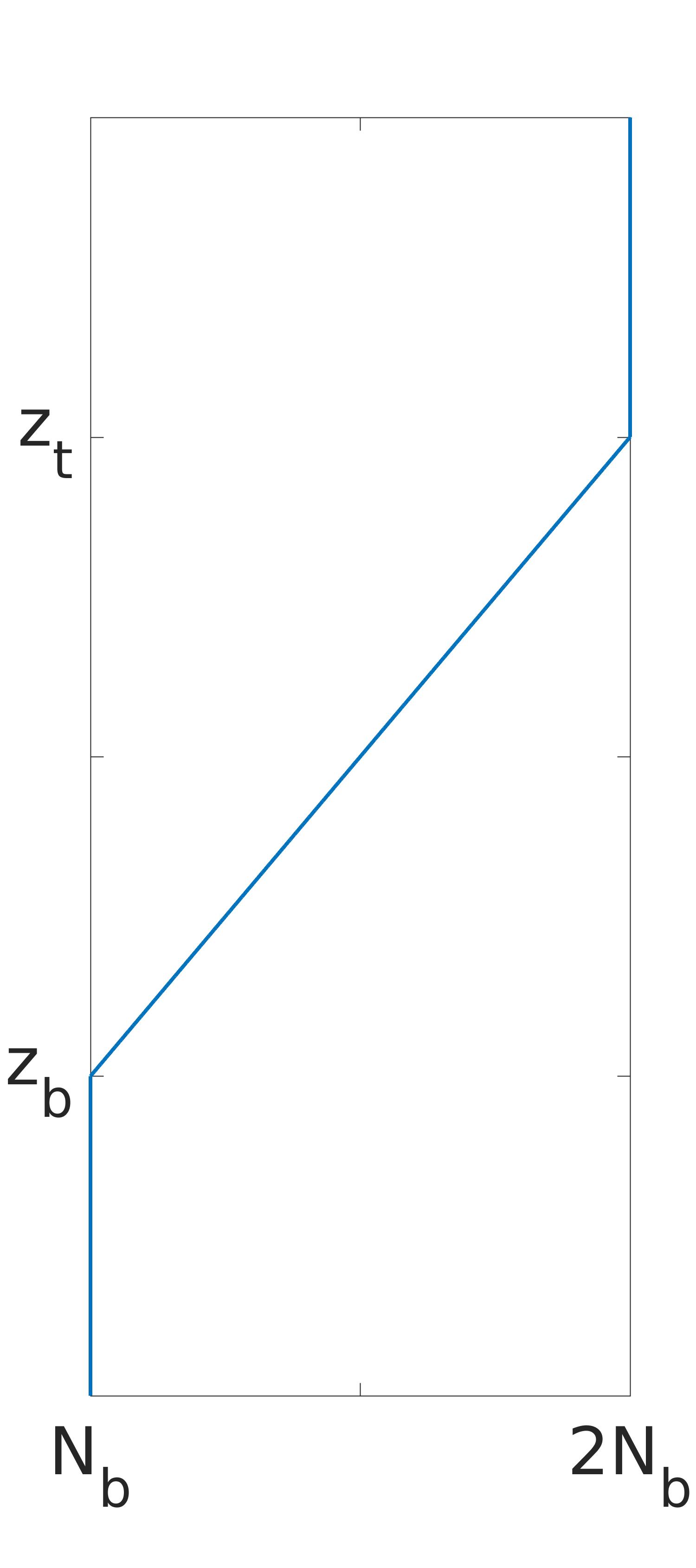}
\includegraphics[width=\textwidth, keepaspectratio]{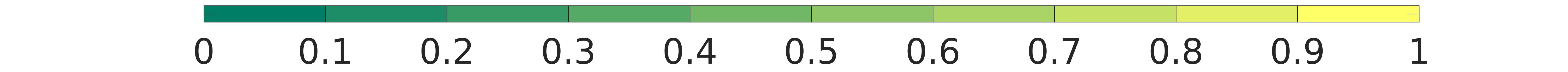}
\caption{{Here we see the results for a stratification profile, that increases linearly over a finite region of depth $\Delta_z$. The right panel is an enlargement of the rightmost part of the left one. The 3 thick lines represent constant vertical wavelength in the bottom layer (it changes while the wave propagates upwards).}}
\label{fig:linear_profile_results}
\end{figure}

The first case for which we will present results is the case of the linearly increasing stratification, as  we defined earlier in equation \eqref{profile:linear_increasing}. We focus on three regimes of horizontal wavelengths: comparable to $\Delta_z $, longer than $\Delta_z$ and shorter than $\Delta_z$. Recall that we are using the Boussinesq approximation, hence $\Delta_z$, as it is also the scale of variation of $N$, is small compared to the density scale height $H_\rho$. So wavelengths that are small and comparable to $\Delta_z$ are also small compared to $H_\rho$. This is the regime where the classical WKB theory is applicable. Ray theory, that is based on WKB assumptions, predicts perfect transmission for those waves in a linearly increasing profile and this is exactly what we are able to find. Even for large horizontal wavelengths, the transmission is high, at least up to a certain point. As we can see in figure \ref{fig:linear_profile_results}, there is stronger reflection when we are moving to the right and to the bottom in the figure, that means that the wave frequency gets closer to the lower value $N_b$ and the horizontal wavelength (and eventually the vertical wavelength) is growing. Figure \ref{fig:linear_profile_results} grants us a closer view into the area of interest, together with some additional information. The solid lines are lines of constant vertical wavelength, which are determined by the gravity wave dispersion relation \eqref{eqn:vertical_wave_number} (they are hyperbolas in the frequency-wave number-space). We see that the vertical wavelength increases by moving to the right and to the bottom in our domain. If the vertical wavelength exceeds $\Delta_z$ by about an order of magnitude, waves start to transmit worse and worse. 

Waves whose wavelengths are small, or at least comparable to the scale of variation in the Brunt-Väisälä-frequency, adapt smoothly to those changes and have a high transmission, while for waves with large wavelengths, the change still happens abruptly. 
Waves that have frequencies close to the Brunt-Väisälä frequency are almost purely horizontal and in the limit of $\omega \to N_b$ for fixed $\lambda_x$, the transmission eventually gets zero. This seems reasonable, since there is no vertical wave structure and hence no vertical energy transport.
In the limit $\lambda_x \to \infty$ (or equivalently $k \to 0$) for fixed $\omega < N_b$, we can find a closed formula for the matrix product in \eqref{eqn:matrix_prod} and hence a formula for the amplitude ratio, which coincides with the classical transmission coefficient of a two-layer model:
\begin{equation}\label{eqn:tc_2layer}
TC_{2L} = \frac{A_{t}}{A_{i}} = \frac{2m_{i}}{m_{i}+m_{t}},
\end{equation}
where the indices describe incident and transmitted wave properties respectively. We adapt the slightly different notation of a transmission coefficient as the ratio of the wave amplitudes due to the classical work that was done in this area, e.g., by \citet{EliassenPalm61}. An analytic execution of these limits can be found in the appendix.


\subsection{Wave tunneling}\label{subsec:tunnelling}

\begin{figure}
\includegraphics[width=0.7544\textwidth, keepaspectratio]{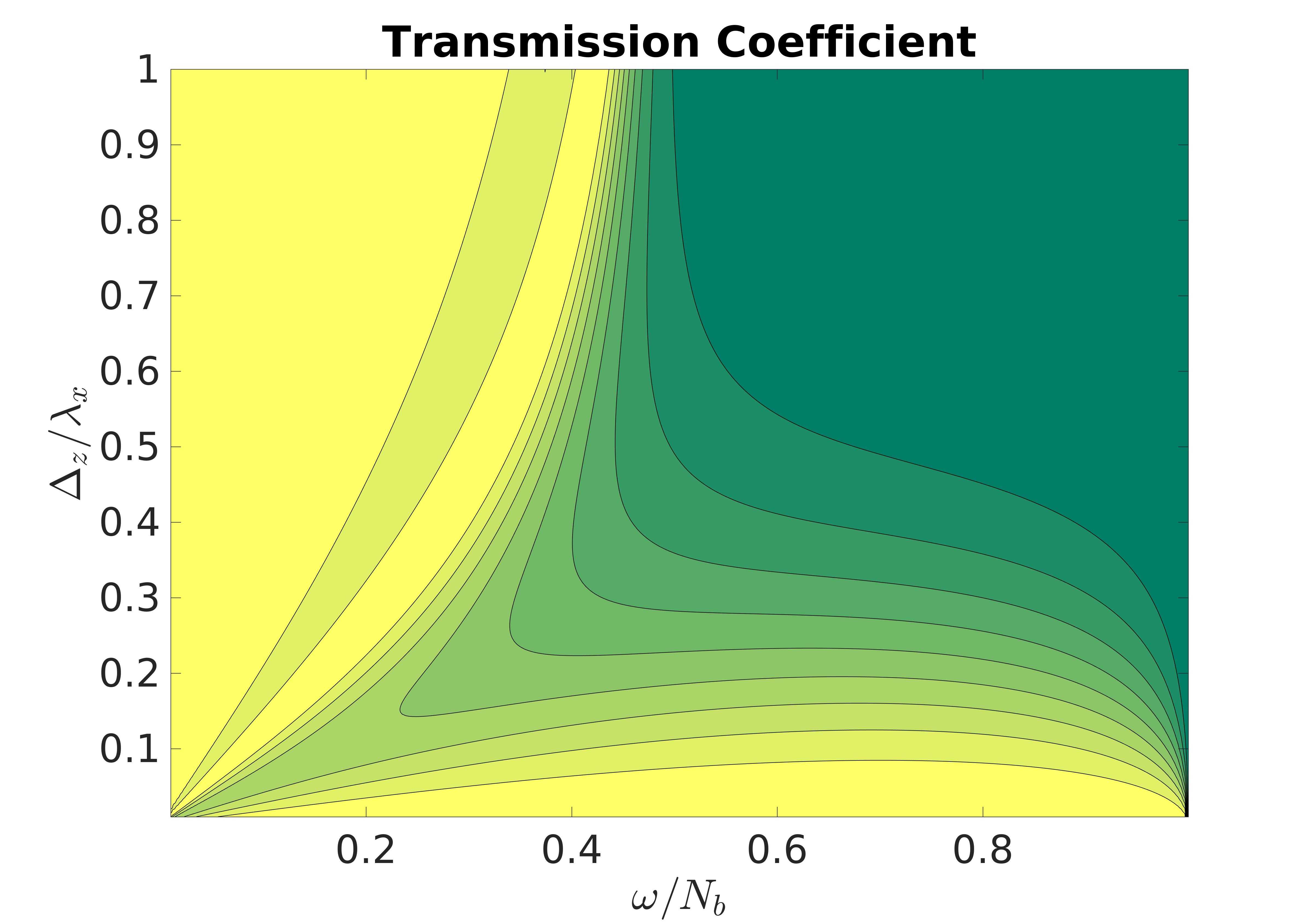}
\includegraphics[width=0.2357\textwidth, keepaspectratio]{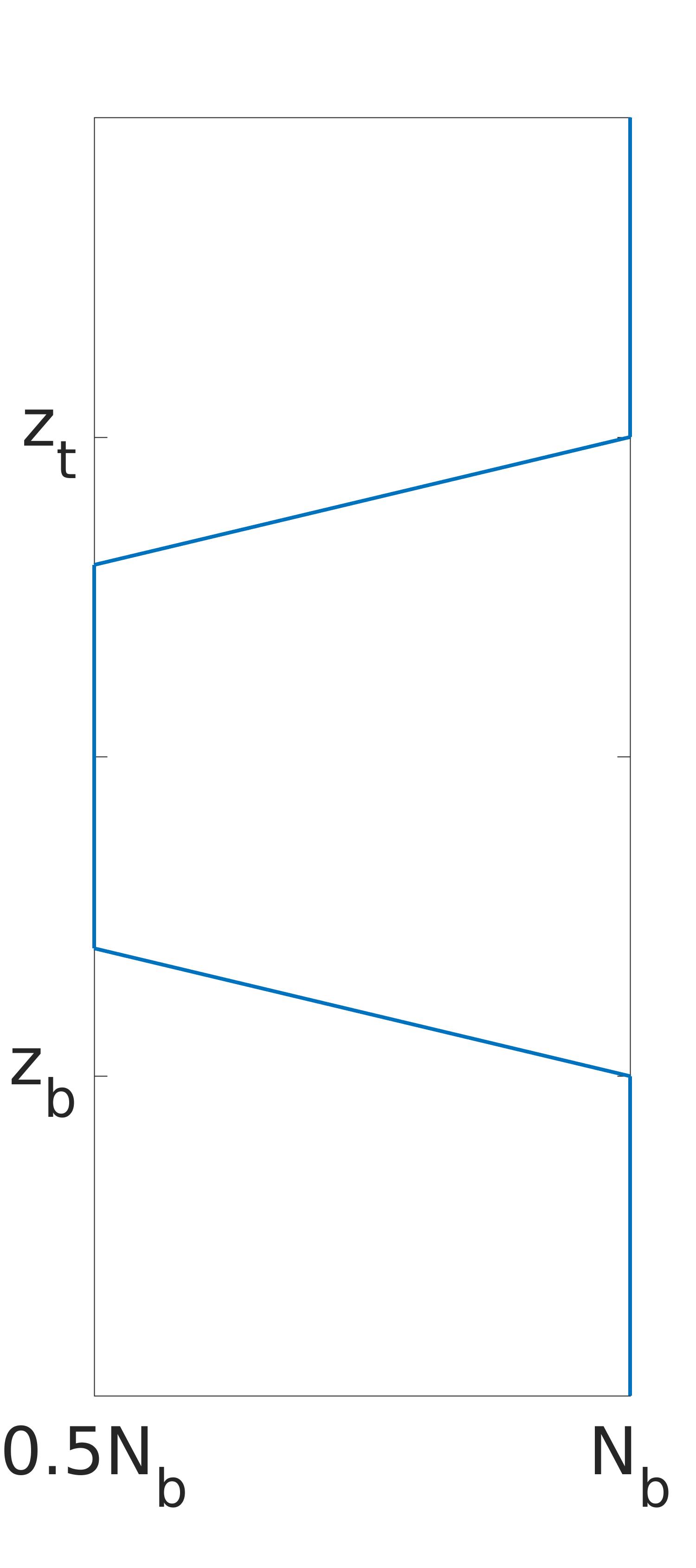}
\includegraphics[width=\textwidth, keepaspectratio]{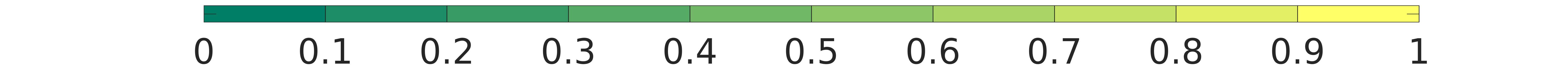}
\caption{The figure shows the transmission coefficients for a profile that has a region of decreased stratification. We see that waves, whose frequency is larger than $0.5 N_b$ can transmit, if their wavelength is long compared to the region of decreased stratification.} 
\label{fig:tunneling}
\end{figure}

We consider now a case where the stratification drops from some $N_b$ to a value $N_d < N_b$ and eventually increases again back to $N_b$:
\begin{equation}\label{profile:decreased}
	N(z) = 
	\begin{cases}
	N_b, & z<z_b \\
	N_b + \frac{z-z_b}{z_{d_1}-z_b}(N_d-N_b), & z_b < z \le z_{d_1}\\	
	N_d, & z_{d_1} \le z \le z_{d_2}\\
	N_d + \frac{z-z_{d_2}}{z_t-z_{d_2}}(N_b-N_d), & z_{d_2} < z \le z_t\\	
	N_b,& z_t < z.
	\end{cases}
\end{equation}
In the example we present, $ z_{d_1}-z_b = 0.2 \Delta_z = z_t - z_{d_2}$ and $N_d = 0.5 N_b$.
The results found for this case are very different from what the classical theory tells us. Ray theory predicts that waves reflect totally from a layer, where $\omega \ge N$. However, when there is only a finite small region where $\omega \ge N$ holds, wave propagation through this region is possible under certain conditions. \citet{SutherlandYewchuck04} described this phenomenon for a sharp drop to a weak or even vanishing stratification, and our results show that tunnelling also exists in the case of a "smooth" transition. In figure \ref{fig:tunneling}, one can see the transmission coefficient for profile \eqref{profile:decreased}. In this particular example, ray theory predicts that every wave with frequency $\omega \ge 0.5 N_b$ would fully reflect from this layer, but we can observe that if the wavelength is large compared to the extent of the region with weak stratification, it is possible to obtain high wave transmission.

As already mentioned in section \ref{sec:convergence}, this case is hard to deal with analytically and numerically. As we will see later in section \ref{sec:numerics}, numerical computations of the full Boussinesq equations show the existence of wave tunnelling (although there are some other difficulties with this case). Moreover, there are lab experiments, for example by \citet{SutherlandYewchuck04}, to further support these findings. {However, the limit ODE found in section \ref{sec:convergence} cannot be solved numerically due to a singularity in the coefficients.} Although there is no evidence that the results are incorrect, the results should mathematically still be taken with a grain of salt.


\subsection{Realistic tropopause profile}

\begin{figure}
\includegraphics[width=0.7544\textwidth, keepaspectratio]{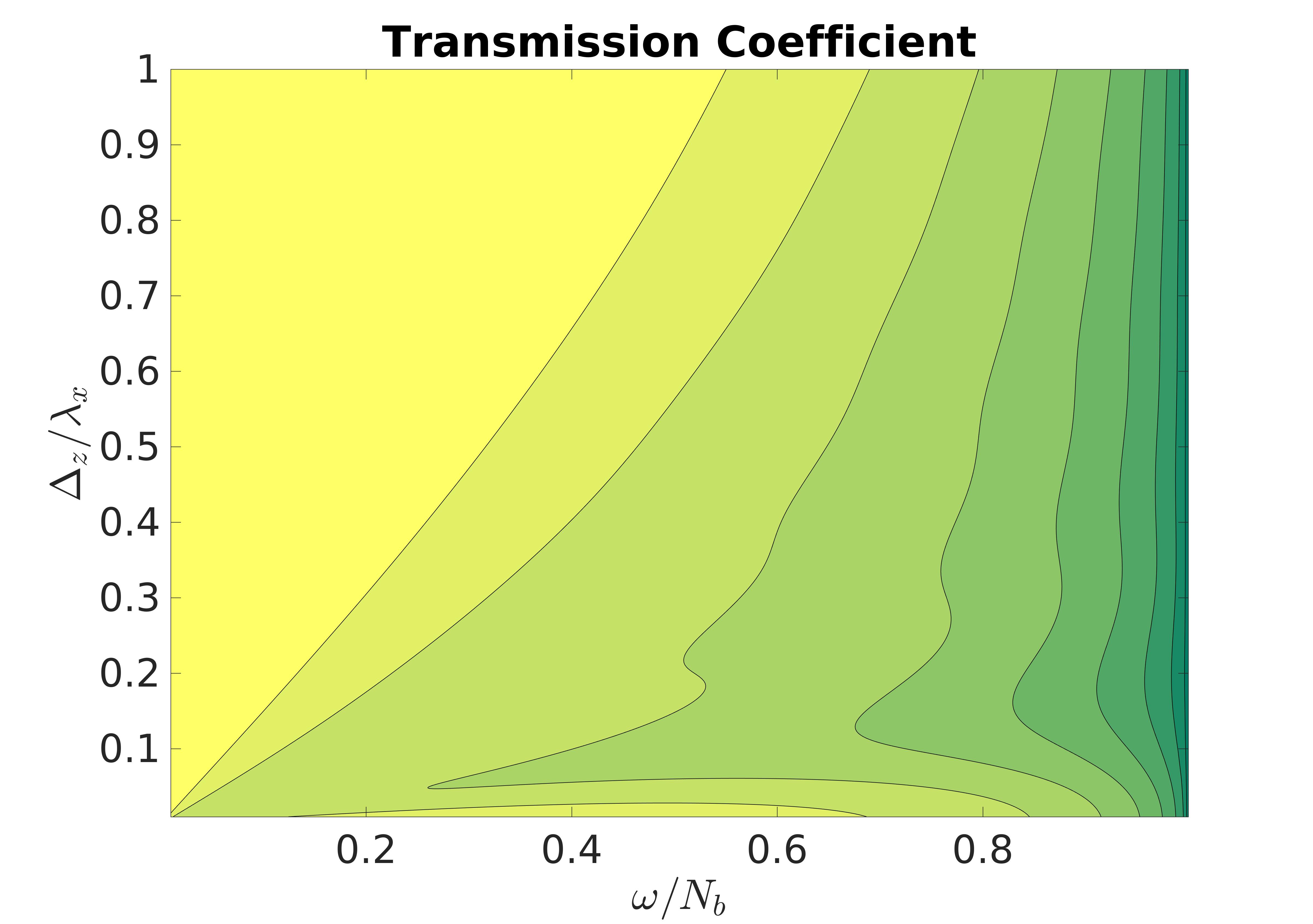}
\includegraphics[width=0.2357\textwidth, keepaspectratio]{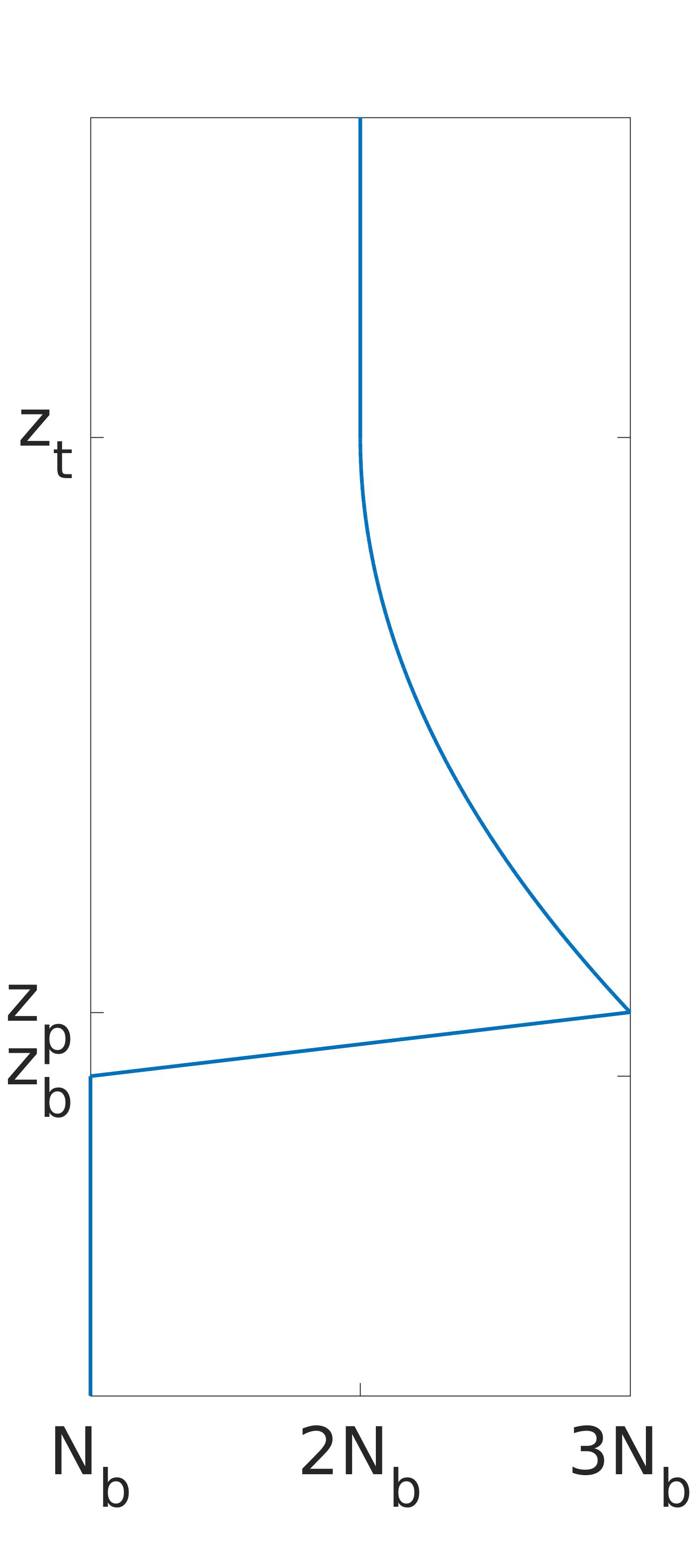}
\includegraphics[width=\textwidth, keepaspectratio]{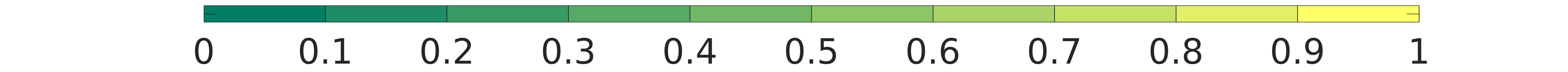}
\caption{Values for the transmission coefficient for a tropopause profile that can bee seen in the right panel. In the limit for long waves, the transmission coefficient approaches again the two-layer solution. For moderately long waves, we can see a combination of 2 effects: The sharp increase, which is almost like a jump and blocks a part of the waves and the smooth relaxation afterwards that has high transmission. We also observe that in the classical WKB regime, the transmission is still very high}
\label{fig:tropopause}
\end{figure}

{Since we are ultimately interested in the behaviour of atmospheric gravity waves and their interaction with the tropopause, we now want to consider a "realistic" tropopause profile}. The stratification is constant with a value $N_b$ below the tropopause. At the temperature inversion layer, the Brunt-Väisälä frequency has a very sharp increase to a peak value $N_p$, almost like a jump, followed by a relaxation to a value $N_t$ with $N_b<N_t<N_p$ that is the constant value of the stratification in the stratosphere. We realize this by a piecewise-defined continuous function:
\begin{equation}
	N(z) = 
	\begin{cases}
	N_b, & z<z_b \\
	N_b + \frac{z-z_b}{z_{p}-z_b}(N_p-N_b), & z_b < z \le z_{p}\\	
	az^2+bz+c, & z_{p} < z \le z_t\\	
	N_t,& z_t < z,
	\end{cases}
\end{equation}
where $a = \frac{N_p-N_t}{(z_p-z_t)^2}, b = -2 z_t \frac{N_p-N_t}{(z_p-z_t)^2}$ and $c = N_t +  z_t^2 \frac{N_p-N_t}{(z_p-z_t)^2}$. The values were chosen such that the profile is continuous at $z_p$ and $z_t$ and differentiable at $z_t$. In the example we show here, the values were set to be $z_p-z_t = 0.1 \Delta_z$, $N_p = 3 N_b$ and $N_t = 2 N_b$. The profile as well as results for the transmission coefficient can be seen in figure \ref{fig:tropopause}. 

For the limits $\omega \to N_b$ for fixed $\lambda_x$ and $\lambda_x \to \infty$ for fixed $\omega$, we have the exact same behaviour as in the linearly increasing case, which is no surprise, since in the first limit, we still have no vertical energy flux and in the second limit we again approach the two-layer model. The rest of the picture however gives some interesting insights. By making the sharp increase asymptotically thin, i.e., making it a (discontinuous) jump, we obtain, for wavelengths comparable to $\Delta_z$, a composition of the transmission coefficient for a two-layer model (that describes the jump) and the one for the smooth profile that follows after the jump. 


\section{Numerical simulations}
\label{sec:numerics}

In this section, results for transmission coefficients retrieved from numerical simulations of 
atmospheric motion are presented, discussed and compared to the results from the multi-layer method
introduced above.

We use two different models to tighten the theoretical findings. The first one is EULAG, a Eulerian/semi-Lagrangian fluid solver described in \citet{Prusa08}, the second one is PincFloit (Pseudo-incompressible flow solver with implicit turbulence model), which was developed by \citet{RieperETAL13}. 


\subsection{Model setup}
\label{subsec:model_setup}

The calculations are done on a two-dimensional $x$-$z$-domain with periodic boundaries
in $x$-direction. Both models, EULAG and PincFloit, are run in Boussinesq mode
and are set up in the following way in order to resemble the 
scenarios for wave transmission discussed above the most.
An absorption layer is located at the bottom of the domain, where the 
flow is relaxed towards a state that is here chosen to be a plane wave field which 
oscillates in space and time.
In order to do this properly, the Brunt-Väisälä frequency needs
to be constant in this sponge layer.
This results in the excitation of a plane wave that can propagate freely
above the sponge layer that covers the lowest 40\% of the domain,
until it 
eventually reaches a region
of non-uniform stratification, the tropopause, which is 
defined as in the three cases discussed in section \ref{sec:results},
where it gets reflected and partially transmitted.
The reflected part gets soaked up by
the bottom sponge while the transmitted part travels in a stratospheric region,
i.e. a region where $N$ is again constant before getting damped away by the
top sponge. This sponge covers the last 20\% of the domain.

The initial wave amplitude has to be chosen
very small, since the theoretical work uses the linearised Boussinesq equations.
Moreover, the waves should not break during the simulations.
Hence we initialise the amplitude as 
\begin{equation}
 b=a\frac{N_b^2}{m_b}
\end{equation}
where $a < 1$ is a parameter denoting to what percentage the threshold of 
static instability $b_{s}= \frac{N_b^2}{m_b}$, found in, e.g., 
\citet{AKS10} is reached.
The quantities of the wave field on the bottom layer can be derived via the
polarisation relations for Boussinesq waves, which can be found in, e.g.,
\citet{AKS10} or \citet{Sutherland10}.
\begin{align}
u&= b \frac{m_b}{k}\,\frac{\omega}{N_b^2} \cos(kx+m_b z-\frac{\pi}{2}-\omega t)\\
w&= b \frac{\omega}{N_b^2}    \cos(kx+m_bz+\frac{\pi}{2}-\omega t)
\end{align}

The top sponge is just an absorption layer, which damps the wave and relaxes to a steady, hydrostatic background. 

The region of non-uniform stratification starts at 60\% of the domain and is in general resolved with 100 grid points. For this value, the theoretical results are very acurate (absolute error is smaller than $10^{-5}$, as seen in section \ref{subsec:error_analysis}). In the reflection layer case, we changed the size of the tropopause rather than the wavelength since we wanted to simulate large ratios between wavelength and tropopause depth, so it was more efficient to shrink the extent of the tropopause.


\subsubsection{EULAG}

The fluid solver EULAG, described by \citet{Prusa08}, 
is used in the Boussinesq framework
(anelastic setup with special choice for the hydrostatic basic state in density $\rho_0$ 
and in potential 
temperature $\theta_0$, namely constant).
The calculations are done on a two-dimensional model domain that extends up to 
$8500\,\mathrm{m}$ in the vertical and has a horizontal width ranging from 
$1000\,\mathrm{m}$ to $3000\,\mathrm{m}$, depending on the wavelength that is prescribed. 
The resolution is chosen to be around $10\,\mathrm{m}$ in $x$ and $z$  
direction.
The vertical extent of the domain is limited by the Boussinesq assumption to a value 
below which the basic state profiles remain physically meaningful.

The transmission coefficient for each time step $t$ is calculated as 
\begin{equation}
 TC(t)=\frac{m_1|A_1|^2}{m_0|A_0|^2}=\frac{m_1\,(\max|w_{ss}(t)|)^2}{m_0\,(\max|w_{ts}(t)|)^2}.
\end{equation}
The transmitted amplitude $A_1$ is determined by finding the maximum absolute vertical wind speed 
in a layer above the region 
of interest (stratosphere: $w_{ss}$) while the amplitude $A_0$ is retrieved from 
a reference simulation with constant stratification and taken as the 
maximum absolute vertical wind speed $w_{ts}$ just below the tropopause altitude.
The depth of these regions is roughly $1\,\mathrm{km}$ for the cases shown here.
This is sufficient since horizontal and vertical
wavelengths are small enough to ensure a maximum or minimum in this box.
Since we expect a static situation for the transmission of a plane wave
through a layer of non-uniform stratification, the simulations are done up to a
simulation time of $15\,\mathrm{h}$ when the simulated flow has stabilized and 
shows little
temporal variation in the quantities of interest. We then take the mean value
of $\overline{TC}$ for the $TC$ calculated at simulation times 
$8\,\mathrm{h},\,9\,\mathrm{h},\,10\,\mathrm{h},\,...,\,15\,\mathrm{h}$.


\subsubsection{PincFloit}

PincFloit has a built-in switch for a Boussinesq atmosphere, i.e. constant background density, but due to extensions implemented by \citet{BoloniETAL16}, it is possible to have a non-constant profile for the Brunt-Väisälä frequency. Hence it was an easy task to implement the test cases of this manuscript into the solver. Moreover, as can be seen for example in \citet{BoloniETAL16} and \citet{SchluKA17}, PincFloit gives very robust results in simulations of atmospheric gravity waves.

The model domain covers 10000\,m in the vertical and one wavelength in the horizontal direction, which lies between 1000\,m and 3000\,m. {The vertical resolution is 10\,m, and the horizontal resolution is 25\,m.}

In all simulated runs, we could observe a stable steady state for several hours of simulation time, which allows us a very good computation of the transmission coefficient. The formula for the computation is, of course, the same as for the EULAG simulations, the choice of incident and transmitted amplitude however is a little bit different. Nonetheless, no method is superior over the other and both yield equally good results. As initial amplitude, we take the mean amplitude of the excited wave in the middle of the bottom sponge layer, as transmitted amplitude we take the mean amplitude of the wave in the stratospheric region, i.e. between 7000 and 8000 m. Moreover, the transmitted amplitude is computed every {10\,min} of simulated time and the transmission coefficient is taken as mean over all computed values {after reaching the steady state. }


\subsection{Results}
\label{ssec:Results}

In this section we show exemplary simulations that cover a portion of the information obtained in
the transmission coefficient figures (see figures \ref{fig:linear_profile_results}, \ref{fig:tunneling}
and \ref{fig:tropopause}) that have been presented before, {as well as snapshots of the wave field after reaching a steady state. This gives a better intuition for wave transmission and reflection. }


\subsubsection{Linear increase}
\label{sssec:LinearIncrease}


\begin{table}
\centering
\begin{tabular}{c|c|c|c|c|c||c}
\diag{0.05em}{0.7cm}{$\lambda_z$}{$\lambda_x$} & 1000 & 1500 & 2000 & 2500 & 3000  & method\\
\hline
\multirow{3}{*}{1000} & 0.9950 & 0.9964 & 0.9979 & 0.9985 & 0.9988 & multi-layer\\
                      & {0.9780} & {0.9713} & {0.9806} & {0.9911} & {0.9972} & PincFloit\\
                      & 0.9727 & 0.9832 & 0.9828 & 0.9881 & 0.9982 & EULAG\\
                      \hline
\multirow{3}{*}{2000} & 0.9560 & 0.9799 & 0.9884 & 0.9892 & 0.9894 & multi-layer \\
                      & {0.9477} & {0.9826} & {0.9948} & {0.9957} & {0.9777} & PincFloit\\
                      & 0.8119 & 0.8390 & 0.8829 & 0.9106 & 0.9297 & EULAG\\             
\end{tabular}
\caption{{Comparison of the transmission coefficients for profile from figure \ref{fig:linear_profile_results} with a dimensional tropopause depth of 1000 m for different horizontal and vertical wavelength (also in m). We can see that there is a very good accordance of the values from the multi-layer method and the ones computed with PincFloit and EULAG simulations. } }
\label{tab:linear}
\end{table}

\begin{figure}
  \centering
   \includegraphics[width=\linewidth]{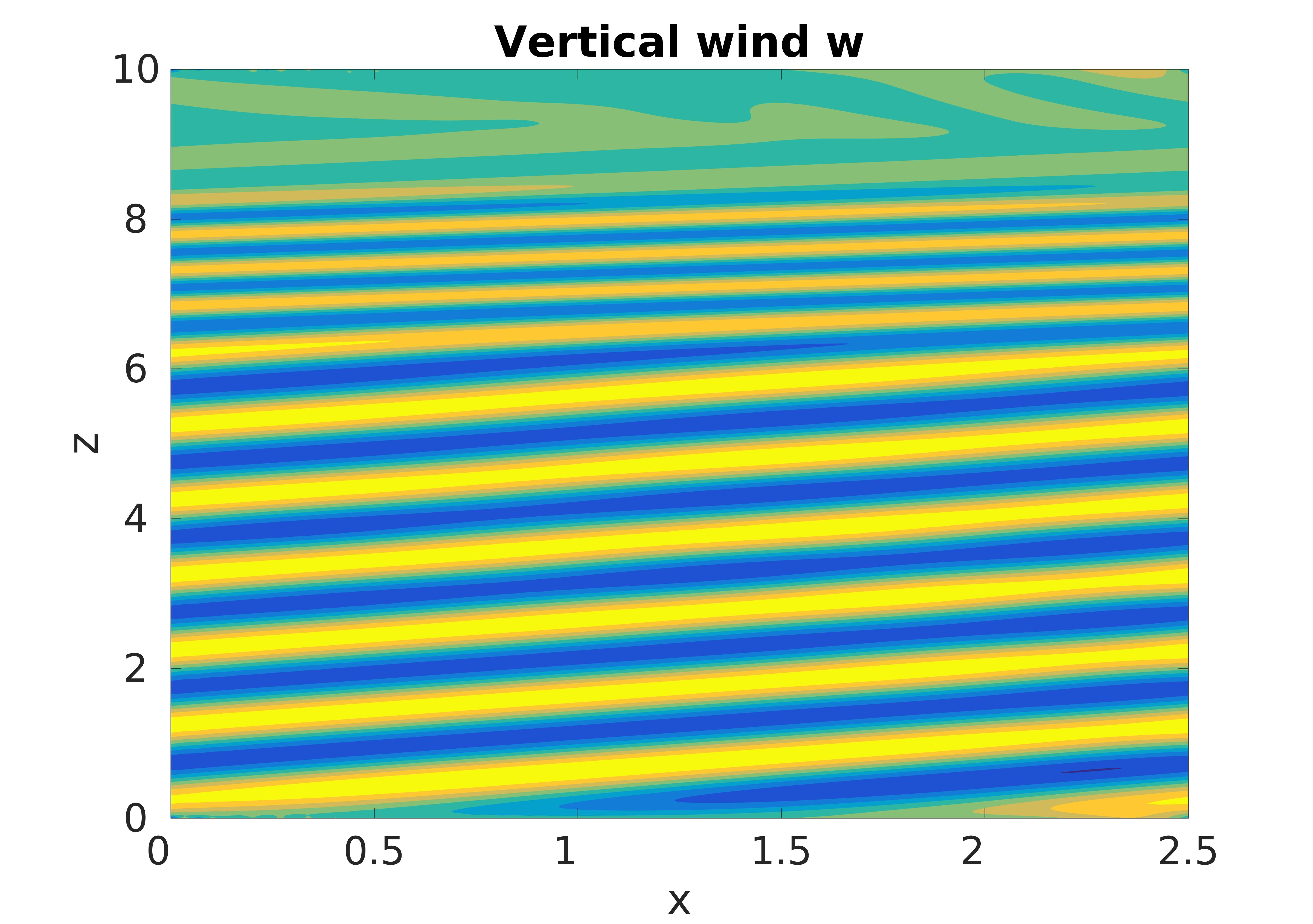}
   \caption{{Vertical wind field of a PincFloit simulation for the linear case with a horizontal wavelength of 2500 m and an incident vertical wavelength of 1000 m. We see no chequerboard pattern, but only a refraction of the wave due to the change in vertical wavelength. This implies that almost no wave reflection occurs. The smaller wave amplitude in the stratosphere (i.e. between 7000 m and 8000 m) is a result of the stronger stratification in this altitude. The axes are normalised by 1000 m.  }}
   \label{fig:pf_linear}
\end{figure}


In table \ref{tab:linear}, we see the comparison of the transmission coefficients obtained from the multi-layer method on one hand and from the numerical simulations of PincFloit and EULAG on the other hand. In general, the values from both methods are in good agreement with the theoretical results.  

Generally speaking, the numerical transmission coefficients are a bit lower than the theoretical ones. This has two reasons. The first one is numerical dissipation which causes a slight decrease of the amplitude over the course of the domain. This is because the wave is initialised in the bottom part of the domain and then freely propagates upward. 
The second reason is concerned with the method we compute incident and reflected amplitude. Due to the nature of the sponge, there are some fluctuations in the initialised wave amplitude and hence also in the transmitted amplitude. We take care of this by averaging over a larger area, but this produces minor errors in the value for the transmission coefficient. 
Plots of the simulated wind field however show that there is little to no reflection at the tropopause in the cases from table \ref{tab:linear}, hence the transmission seems to be even closer to the theoretically predicted values.


\subsubsection{Wave tunnelling}
\label{sssec:WaveTunneling}

%
\begin{table}
\centering
\begin{tabular}{c|c|c|c|c||c}
$\frac{\Delta_z}{\lambda_z}$ & 0.1 & 0.2 & 0.5 & 1 & method\\
\hline
\multirow{3}{*}{TC} & 0.8648 & 0.5846 & 0.0916 & 0.0028 & multi-layer\\
                    & 0.8180 & 0.4985 & 0.0971 & 0.0030 & PincFloit\\
                    & 0.8732 & 0.5659 & 0.1137 & 0.0521 & EULAG\\
\end{tabular}
\caption{{Comparison of the transmission coefficients for profile from figure \ref{fig:tunneling} with fixed vertical and horizontal wavelengths $\lambda_z = \lambda_x$ and varying tropopause depth $\Delta_z$. This corresponds to a vertical cut along $\omega \approx 0.7$ in the left panel of figure \ref{fig:tunneling}. As one would expect from the theory, waves almost fully reflect, when the tropopause is comparable in size to their wavelength, but they transmit very good for a thin tropopause. Both models are able to capture this phenomenon very accurately.} }
\label{tab:reflection}
\end{table}
\begin{figure}
\includegraphics[width=0.497\textwidth, keepaspectratio]{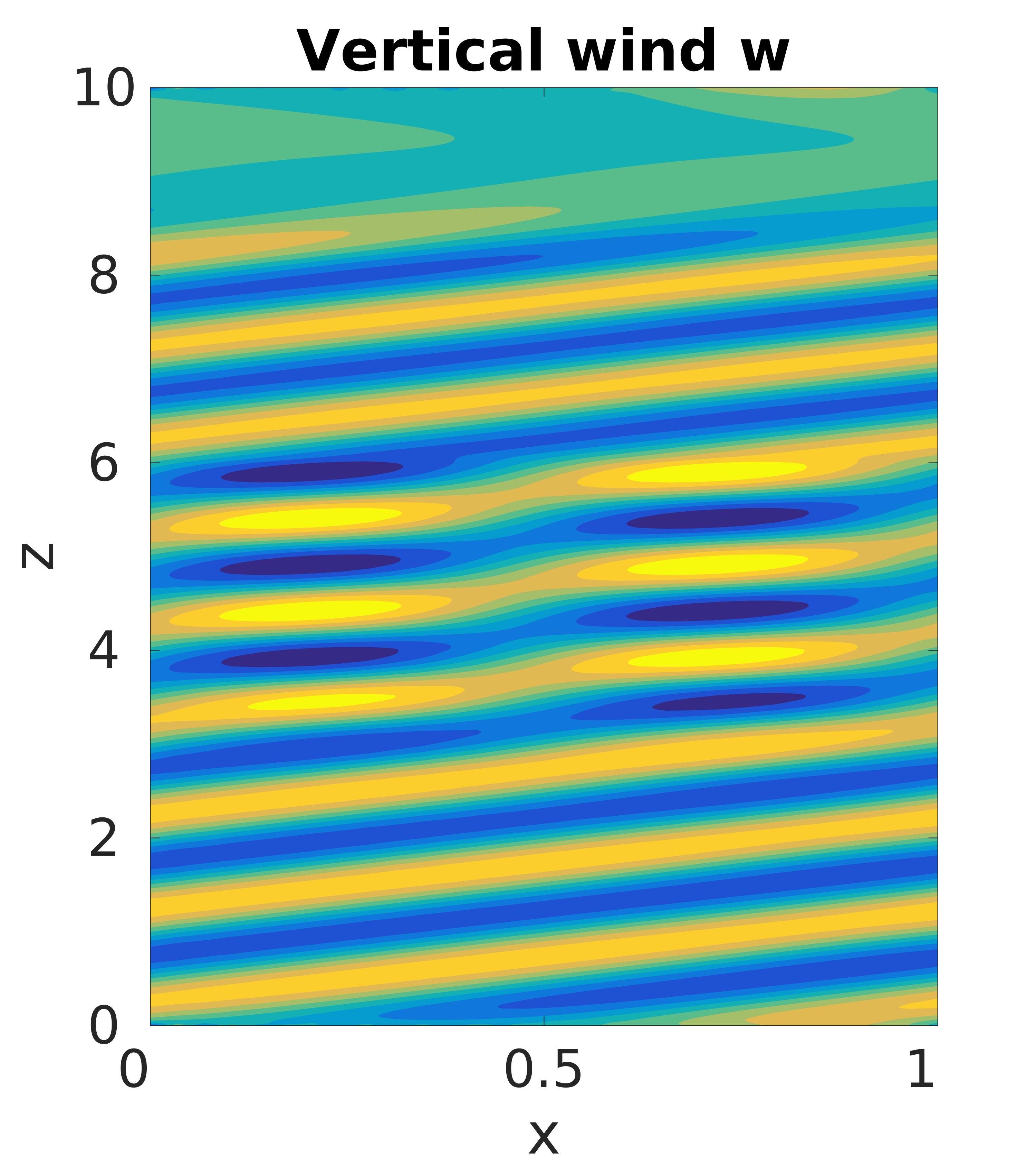}
\includegraphics[width=0.497\textwidth, keepaspectratio]{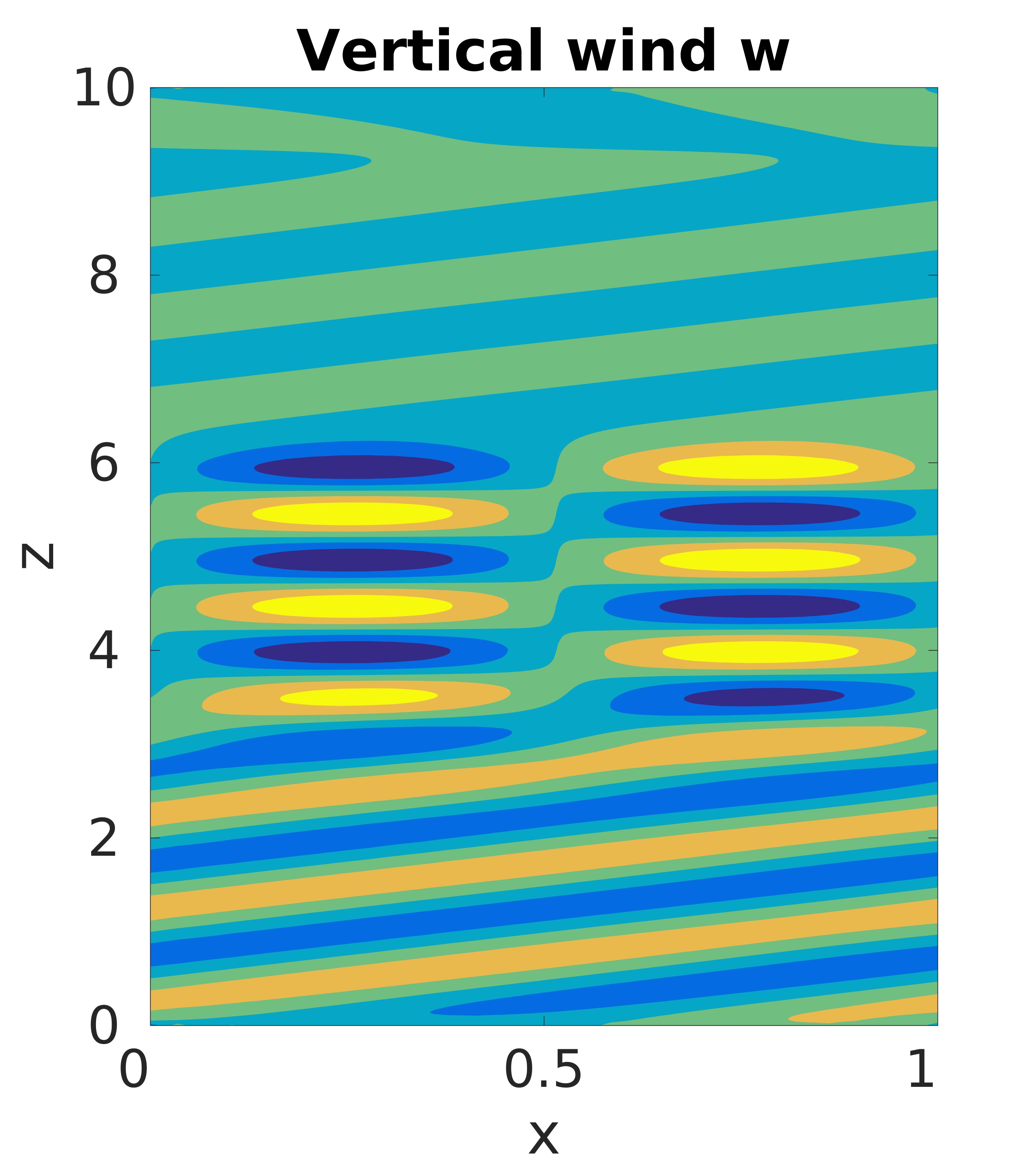}
\caption{ {Snapshots of the wave field after 8 hours of simulated time from PincFloit simulations for the profile from figure \ref{fig:tunneling}. In both cases, we have horizontal and vertical wavelengths of 1000 m. In the left panel, the tropopause depth is 100 m. There is only a slight alternation visible below the tropopause and comparable amplitudes can be observed above and below the tropopause. This suggests a high transmission. In the right panel, the tropopause depth is 500 m. A strongly alternating pattern can be found below the tropopause. Together with the low the amplitudes in the stratosphere, this signifies that most of the wave is reflected. In both panels, the axes are normalised by 1000 m. } }
\label{fig:wind_tunnelling}
\end{figure} 

{The simulation setup in this case uses fixed values for horizontal and vertical wavelength, but changes the depth of the tropopause, as this is numerically more convenient. Technically, we have the same wave frequency, but change the ratio between horizontal/vertical wavelength and tropopause depth, so this corresponds to a vertical slice in the left panel of figure \ref{fig:tunneling}. By choosing $\lambda_x=\lambda_z$, as we did here, this results in a frequency $\omega=\frac{1}{\sqrt{2}} N_b \approx 0.707 N_b$. For tropopause depths that are comparable to $\lambda_x$, we expect strong reflection, while for a very short tropopause, we should obtain a wave tunnelling effect. The results can be seen in table \ref{tab:reflection}. The PincFloit simulations used $\lambda_x=\lambda_z=1000 \mathrm{m}$, while EULAG used $\lambda_x=\lambda_z=2000 \mathrm{m}$.

Both models match the theoretical prediction very accurately. While the transmission is very high for a short tropopause, we get almost full reflection when it grows in size. Snapshots of the steady state for a high transmission and low transmission case can be seen in figure \ref{fig:wind_tunnelling}. It gives a very good comparison between the two cases as we can clearly see the differently pronounced alternating patterns as well as the disparity in the stratospheric amplitudes. 
}


\subsubsection{Realistic tropopause profile}
\label{sssec:RealisticTropopauseProfile}


\begin{table}[h]
\centering
\begin{tabular}{c|c|c|c|c|c||c}
\diag{0.05em}{0.7cm}{$\lambda_z$}{$\lambda_x$} & 1000 & 1500 & 2000 & 2500 & 3000 & method\\
\hline
\multirow{3}{*}{1000} & 0.7858 & 0.8010 & 0.8095 & 0.8151 & 0.8185 & multi-layer\\
                      & 0.7723 & 0.7914 & 0.7805 & 0.8320 & 0.8262 & PincFloit\\
                      & 0.7884 & 0.8084 & 0.8169 & 0.8224 & 0.8367 & EULAG\\
                      \hline
\multirow{3}{*}{2000} & 0.5635 & 0.6237 & 0.6620 & 0.6913 & 0.7113 & multi-layer \\
                      & 0.5375 & 0.6182 & 0.6600 & 0.6763 & 0.7204 & PincFloit \\
                      & 0.5559 & 0.5912 & 0.6248 & 0.6565 & 0.6814 & EULAG\\
\end{tabular}
\caption{{Comparison of the transmission coefficients for profile from figure \ref{fig:tropopause} with a dimensional tropopause depth of 1000 m for different horizontal and vertical wavelength (also in m). We can see that there is a very good accordance of the values from the multi-layer method and the ones computed from the simulations.  } }
\label{tab:tropopause}
\end{table}
\begin{figure}[t]
  \centering
   \includegraphics[width=\linewidth]{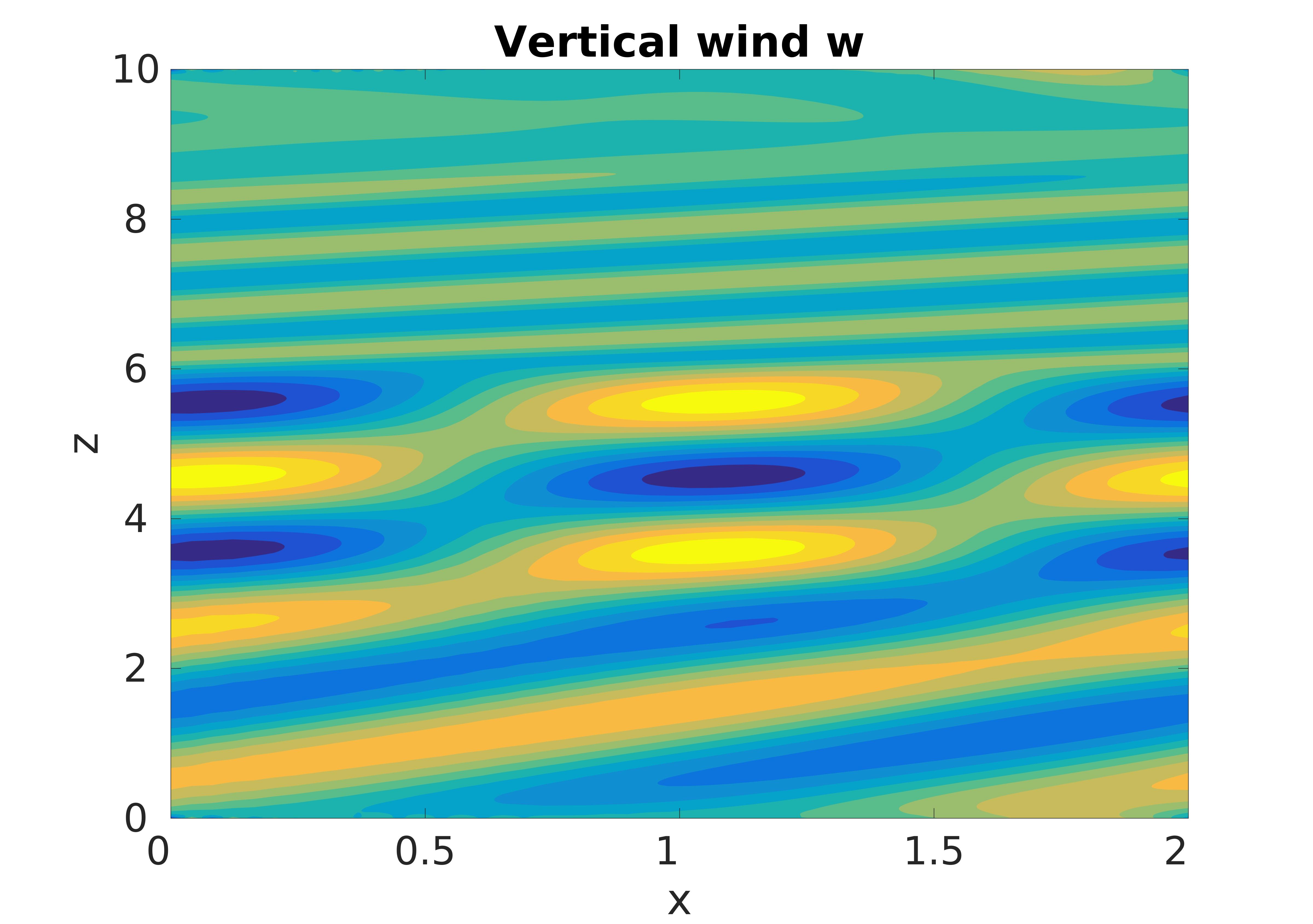}
   \caption{{Vertical wind field of a PincFloit simulation for the case of the realistic tropopause profile with a horizontal wavelength of 2000 m and a vertical wavelength of 2000 m and with a tropopause depth of \mbox{1000 m}. We see a moderately alternating pattern and waves above the tropopause. This indicates partial wave reflection and transmission. The computed transmission coefficient is 0.6600. It is also possible to see the work of the bottom sponge layer, which extends from 0 m to \mbox{4000 m} in the vertical direction. The superposition of upward and downward propagating wave is still clearly visible in the upper part of the sponge while they are completely damped away in the bottom part. The axes are normalised by $1000\,\mathrm{m}$.}}
   \label{fig:wind_tp}
\end{figure}

Table \ref{tab:tropopause} shows a comparison of values for the transmission coefficient obtained from the multi-layer method and values computed from PincFloit and EULAG simulations. As in the two other cases, we can see a very good agreement between theory and numerics. These cases were particularly interesting since there is only a partial reflection of the wave. Both codes were able to reproduce this phenomenon. A snapshot of how a wave field looks like in this case can be seen in figure \ref{fig:wind_tp}.


\subsection{Conclusions}
\label{ssec:Conclusions}

The simulations deliver very accurate results. The qualitative
as well as the quantitative behaviour could be reproduced in every single of
the atmospheric test cases with different parameters. Even wave tunnelling
could be observed. This leads to the conclusion that the multi-layer method
is capable of predicting the transmission and reflection of gravity waves on a
satisfying level. Since the computation is very fast and efficient, this could find
use as a black box in numerical weather models, whose resolution is in general
too coarse to resolve all gravity wave structures. The prediction of how much
of the wave energy transmits through the tropopause can lead to a better
parametrisation of gravity waves in the middle and upper atmosphere, and
could also explain why some of the waves that are supposed to have broken at
a certain height are still present and stable. Moreover, stronger temperature
inversions in the tropopause lead to larger wave reflection with a corresponding
downdraught of energy which could affect the tropospheric balance.

\section{Further discussion}
\label{sec:FurtherDiscussion}

This section is dedicated to limitations as well as the possible extensions of the model. Since we are using the Boussinesq approximation, there are clearly some bounds on the applicability in case the fluid density is changing significantly. But there is also room for improvement and generalisation. Background wind is an important factor in the atmosphere as well as wave packets, i.e. locally confined wave movements whose envelope is moving with the group velocity.


\subsection{Limitations}
\label{ssec:Limitations}

Waves with a frequency higher than the Brunt-Väisälä frequency are evanescent. However, we have seen in Section \ref{subsec:tunnelling} that if the region over which the waves are evanescent is small compared to the wavelengths, the waves can survive and eventually propagate again. Hence one would expect a similar behaviour if waves are excited with a frequency higher than the Brunt-Väisälä frequency, but after a short extent reach a region where it gets and stays larger than their own frequency. Although there is nothing that speaks against this hypothesis, our approach as it is presented here, is not capable of reproducing this effect. We are assuming an infinitely extended domain of constant stratification above and below the non-uniform region and that incident waves come from "far away".  Hence a wave that is initially evanescent will completely vanish (i.e.\ amplitude $A=0$) when reaching the non-uniform region. 

A workaround for this would be to assume a stratification profile that has an infinite bottom layer that allows for wave propagation with a certain frequency $\omega_0$, followed by a jump to a confined region where the Brunt-Väisälä frequency is smaller than $\omega_0$, which in turn is succeeded by the profile we are interested in. The relevant amplitude for the transmission coefficient computation would then be the upward amplitude in the confined region of reduced stratification. This is possible since the multi-layer method keeps track of upward and downward propagating wave amplitudes.


\subsection{Extensions}
\label{ssec:Extensions}


\subsubsection*{Including background wind}

All previous computations are done for an atmosphere at rest. The case with background wind can be more difficult, since not wave energy, but rather wave action is conserved. However, since we are interested in the wave action fluxes, the formula for the transmission coefficient does not change. We have to distinguish two different cases: constant and non-constant background wind.

In the case of constant background wind $U =U_0 \equiv  \text{const.}$, the results are pretty much the same as with no background wind, except that one has to use the intrinsic frequency $\hat \omega = \omega - k U_0$ instead of the extrinsic frequency $\omega$ (which are in fact equal if there is no background wind). Waves can then propagate for $\hat \omega \in \left( 0,N_b \right)$.  {Apart from that small change, the multi-layer method, as it is presented here, can be applied without further modification.}

If the background wind is changing with height, it will have an impact on the vertical propagation of the gravity waves just as much as the non-uniform stratification has. In this case, we have to take into account the change in absolute frequency over the shear region as well as the curvature of the mean background wind, which is the rate of change of the shear. Moreover, we can have reflection layers, similar to the case of weakening stratification, and also a new phenomenon, called critical layers, arises. They occur at locations where the absolute frequency vanishes, i.e., when the horizontal phase speed matches the background wind speed. Around critical layers, non-linear effects become more and more important. For example, wave-mean flow interaction, where the wave deposits energy to or draws it from the mean flow, is an important factor that our approach does not account for. But apart from critical layers, which are also an issue in direct numerical simulations, the implementation of non-constant background wind is possible, but will be postponed  {to a companion paper \cite{PuetzKlein2018}}.


\subsubsection*{Wave packets}

Wave packets are an important part of atmospheric gravity wave analysis, since a long-lasting source of  wave generation, such as steady flow over a mountain ridge, is a rare event. A wave packet can be seen as an amplitude-modulated plane wave, with almost compact support, i.e., the envelope vanishes or approaches zero outside a specific region. Mathematically, this can be described as the superposition of infinitely many plane waves with different wave numbers. They destructively interfere almost everywhere except for a confined region in which the interference is constructive. Fourier transformation can be used to break the wave packet down into the different wave numbers with respective amplitudes and phase shifts.

It is possible to apply the method introduced in this paper to wave packets by approximating them as a superposition of finitely many plane waves with corresponding amplitudes. Since it is not only about applying the multi-layer method for some selected wave parameters, but also requires some additional set-up, we will not include the details here. {Alongside the inclusion of background wind, this will be the main issue of a companion paper \cite{PuetzKlein2018}.} 


\section{Conclusions}
\label{sec:Conclusions}

We developed a method with which we can compute the transmission of gravity waves through a finite region of non-uniform stratification by modelling this region as a multi-layer fluid which has a uniform stratification in each layer. The solutions we found for each layer were matched across the fluid interfaces and we were able to relate incident and transmitted wave. The method is applicable to any stratification profile. Moreover, it is able to keep track of the upward and downward propagating wave amplitude at any point in the domain.

We were also able to find the limit for the number of layers tending to infinity, leading us to a system of ordinary differential equations for the amplitudes of upward and downward propagating wave. For stratification profiles not including a reflection layer, this ODE system could be solved numerically in a very efficient way by using a basic integration scheme. An error analysis showed that the multi-layer method converges to the limit solution. 

Numerical simulations of the full Boussinesq equations were carried out to support our theoretical findings. We found the results of the simulations to be in good agreement to the predictions from the multi-layer method.

\section*{Acknowledgements}
The data for this paper are available upon request from the authors. The authors thank the German Research Foundation (DFG) for support through the research unit Multi-Scale Dynamics of Gravity Waves (MS-GWaves) and through grants KL611/24-1, KL611/25-1 and SP1163/5-1. They also thank the group of Prof.\ Ulrich Achatz for the provision of PincFloit.

\newpage

\begin{thebibliography}{19}
\providecommand{\natexlab}[1]{#1}
\providecommand{\url}[1]{\texttt{#1}}
\expandafter\ifx\csname urlstyle\endcsname\relax
  \providecommand{\doi}[1]{doi: #1}\else
  \providecommand{\doi}{doi: \begingroup \urlstyle{rm}\Url}\fi

\bibitem[Achatz et~al.(2010)Achatz, Klein, and Senf]{AKS10}
Ulrich Achatz, Rupert Klein, and Fabian Senf.
\newblock Gravity waves, scale asymptotics and the pseudo-incompressible
  equations.
\newblock \emph{J. Fluid Mech.}, 2010.

\bibitem[Booker and Bretherton(1967)]{BookerBretherton67}
John~R. Booker and Francis~P. Bretherton.
\newblock The critical layer for internal gravity waves in a shear flow.
\newblock \emph{J. Fluid Mech.}, 1967.

\bibitem[Brown and Sutherland(2007)]{BrownSutherland06}
G.~L. Brown and Bruce~R. Sutherland.
\newblock Internal wave tunnelling through non-uniformly stratified shear flow.
\newblock \emph{Atmosphere-Ocean}, \textbf{45}, 2007.

\bibitem[Bölöni et~al.(2016)Bölöni, RIbstein, Muraschko, Sgoff, Wei, and
  Achatz]{BoloniETAL16}
Gergely Bölöni, Bruno RIbstein, Jewgenija Muraschko, Christine Sgoff, Junhong
  Wei, and Ulrich Achatz.
\newblock The interaction between atmospheric gravity waves and large scale
  flows: an efficient description beyond the nonacceleration paradigm.
\newblock \emph{J. Atmos. Sci.}, 2016.

\bibitem[Bühler(2009)]{Buehler09}
Oliver Bühler.
\newblock \emph{Waves and mean flows}.
\newblock Cambridge University Press, 2009.

\bibitem[Danielsen and Bleck(1970)]{DanielsenBleck70}
Edwin~F. Danielsen and Rainer Bleck.
\newblock Tropospheric and stratospheric ducting of stationary mountain lee
  waves.
\newblock \emph{J. Atmos. Sci.}, \textbf{27}, 1970.

\bibitem[Drazin and Reid(1981)]{DrazinReid81}
P.~G. Drazin and W.~H. Reid.
\newblock \emph{Hydrodynamic Stability}.
\newblock Cambridge University Press, 1981.

\bibitem[Eliassen and Palm(1961)]{EliassenPalm61}
Arndt Eliassen and Enok Palm.
\newblock On the transfer of energy in stationary mountain waves.
\newblock \emph{Geofys. Publ.}, \textbf{22}:\penalty0 1--23, 1961.

\bibitem[Lara(2004)]{Lara04}
Luis Lara.
\newblock A numerical method for solving a system of nonautonomous linear
  ordinary differential equations.
\newblock \emph{Appl. Math. Comput.}, 2004.

\bibitem[Mercier et~al.(2008)Mercier, Garnier, and Dauxois]{MercierETAL08}
Matthieu~J. Mercier, Nicolas~B. Garnier, and Thierry Dauxois.
\newblock Reflection and diffraction of internal waves analyzed by the hilbert
  transform.
\newblock \emph{Phys. of fluids}, 20, 2008.

\bibitem[Nault and Sutherland(2007)]{NaultSutherland07}
J.~T. Nault and Bruce~R. Sutherland.
\newblock Internal wave transmission in nonuniform flows.
\newblock \emph{Phys. Fluids}, \textbf{19}\penalty0 (016601), 2007.

\bibitem[Prusa et~al.(2008)Prusa, Smolarkiewicz, and Wyszogrodzki]{Prusa08}
J.~M. Prusa, P.~K. Smolarkiewicz, and A.~A. Wyszogrodzki.
\newblock Eulag, a compuational model for multiscale flows.
\newblock \emph{J. Comput. Fluids}, 2008.

\bibitem[Rieper et~al.(2013)Rieper, Hickel, and Achatz]{RieperETAL13}
Felix Rieper, Stefan Hickel, and Ulrich Achatz.
\newblock A conservative integration of the pseudo-incompressible equations
  with implicit turbulence parametrization.
\newblock \emph{Mon. Weather Rev.}, 2013.

\bibitem[Schlutow et~al.(2017)Schlutow, Klein, and Achatz]{SchluKA17}
Mark Schlutow, Rupert Klein, and Ulrich Achatz.
\newblock Finite-amplitude gravity waves in the atmosphere: travelling wave
  solutions.
\newblock \emph{J. Fluid Mech.}, 826, 2017.

\bibitem[Scorer(1949)]{Scorer49}
Richard~S. Scorer.
\newblock The theory of waves in the lee of mountains.
\newblock \emph{Quart. J. R. Met. Soc.}, \textbf{75}, 1949.

\bibitem[Stalker(1998)]{Stalker98}
John Stalker.
\newblock \emph{Complex Analysis}.
\newblock Birkhäuser, 1998.

\bibitem[Sutherland(2010)]{Sutherland10}
Bruce~R. Sutherland.
\newblock \emph{Internal Gravity Waves}.
\newblock Cambridge University Press, 2010.

\bibitem[Sutherland and Yewchuck(2004)]{SutherlandYewchuck04}
Bruce~R. Sutherland and Kerianne Yewchuck.
\newblock Internal wave tunnelling.
\newblock \emph{J. Fluid Mech.}, 2004.

\bibitem[Teschl(2012)]{Teschl12}
Gerald Teschl.
\newblock \emph{Ordinary Differential Equations and Dynamical Systems}, volume
  140 of \emph{Graduate studies in mathematics}.
\newblock American Mathematical Society, 2012.

\bibitem[Puetz and Klein (2018)]{PuetzKlein2018}
Christopher Pütz and Rupert Klein.
\newblock Initiation of ray tracing models: Evolution of small-amplitude gravity wave packets in non-uniform background.
\newblock \emph{Theor. Comp. Fluid Dyn.}, submitted.
\end{thebibliography}


\newpage

\section*{Appendix A}
Here is the proper execution of the limit process mentioned in equation \eqref{eqn:limit_matrix_diagonal_entry}:
\begin{equation}
\begin{aligned}
{c_j-1}& = {\frac{1}{2} \left(  \frac{m(z_{j})}{m(z_{j+1})}+1 \right) \exp\left(i(m(z_j)-m(z_{j+1}))z_j\right) -1 }{}\\
&={\frac{1}{2} \left(  \frac{m(z_{j})}{m(z_{j}+h)}+1 \right) \exp\left(i(m(z_j)-m(z_{j}+h))z_j\right) -1 }\\
&= \frac{1}{2} \left(\frac{m(z_j)}{m(z_j) + hm^\prime(z_j)+o(h)} + 1\right) \\
& \quad \cdot \exp(ih(m^\prime(z_j)+o(h))z_j)   -1\\
&= \frac{1}{2} \left(\frac{m(z_j)}{m(z_j) + hm^\prime(z_j)+o(h)} + 1\right) \\
& \quad \cdot \left(1+ihm^\prime(z_j)z_j+o(h)\right)   -1\\
&= {\frac{1}{2} \left(\frac{2m(z_j) + hm^\prime(z_j)}{m(z_j) + hm^\prime(z_j)+o(h)}\right)\left(1-ihm^\prime(z_j)z_j+o(h)\right)   -1}{}\\
&= { \left(\frac{- hm^\prime(z_j)}{2(m(z_j) + hm^\prime(z_j)+o(h))}\right)  }{} \\
& \quad - h\left(\frac{2m(z_j) + hm^\prime(z_j)}{2(m(z_j) + hm^\prime(z_j)+o(h))}\right) im^\prime(z_j)z_j+o(h)  \\
&= h\left(\frac{-m^\prime(z_j)}{2(m(z_j) + hm^\prime(z_j)+o(h))}\right)  \\
& \quad - h\left(\frac{2m(z_j) + hm^\prime(z_j)}{2(m(z_j) + hm^\prime(z_j)+o(h))}\right) im^\prime(z_j)z_j+o(h).
\end{aligned}
\end{equation}
We are able to compute the limit of $\frac{c_j-1}{h}$
\begin{equation}
f(z) \wdef \lim_{h \to 0} \frac{c_j-1}{h} = -\frac{m^\prime(z)}{2m(z)} -im^\prime(z)z.
\end{equation}
A similar derivation can be made for $\frac{d_j}{h}$. 

\section*{Appendix B}

We investigate the limit $\omega \to N_b$ for fixed $k$ and the limit $k \to 0$ for fixed $\omega$. For the first limit, we have a look at the following: 
\begin{equation}
\lim_{\omega \to N_b} m_j = \lim_{\omega \to N_b} -k \sqrt{\frac{N_j^2}{\omega^2}-1} = -k \sqrt{\frac{N_j^2}{N_b^2}-1} \wdefi \tilde m_j,
\end{equation}
If $j \neq 1$, then $N_j \neq N_b$ (remember: linear increasing profile). Hence, $\tilde m_j \neq 0$. Since $\det \mathbold{M}_j = \frac{m_j}{m_{j+1}}$, $\mathbold{M}_j$ is regular for $J \neq 1$. For $j=1$, $N_1 = N_b$, hence $\tilde m_1 = 0$ and therefore $\det \mathbold M_1 = 0$. Since the determinant is multiplicative, the determinant of the matrix product \eqref{eqn:matrix_prod} is 0, and by the definition of the transmission coefficient \eqref{eqn:TC}, the transmission is also 0. 

For the second limit, we write
\begin{equation}
 m_j = k \sqrt{\frac{N_j^2}{\omega^2}-1} = k \hat m_j
\end{equation}
and consider the matrix entries \eqref{eqnsys:matrix_entires} in the limit $k \to 0$:
\begin{equation}
\begin{aligned}
\tilde c_j \wdef\lim_{k \to 0} c_j & = \lim_{k \to 0} \frac{1}{2}\left(\frac{m_j}{m_{j+1}}+1 \right) \exp \left( i\left(m_j-m_{j+1}z_j\right)\right)\\
&= \lim_{k \to 0} \frac{1}{2}\left(\frac{k \hat m_j}{k \hat m_{j+1}}+1 \right) \exp \left( ik\left(\hat m_j-\hat m_{j+1}z_j\right)\right)\\
&= \lim_{k \to 0} \frac{1}{2}\left(\frac{ \hat m_j}{ \hat m_{j+1}}+1 \right) \left(1 + ik\left(\hat m_j-\hat m_{j+1}z_j\right) + o(k) \right)\\
&= \frac{1}{2}\left(\frac{ \hat m_j}{ \hat m_{j+1}}+1 \right)
\end{aligned}
\end{equation}
In a similar fashion, we follow that 
\begin{align}
\tilde d_j \wdef \lim_{k \to 0}d_j &= \frac{1}{2}\left(\frac{ \hat m_j}{ \hat m_{j+1}}-1 \right).
\end{align}
If we assume now, that there is no reflection layer, i.e. $m_j \in \R$ for all $j$, then the matrices $\widetilde{\mathbold{M}}_j \wdef \lim_{k \to 0} \mathbold{M}_j$ are real and symmetric. Hence the product of any two matrices of this kind is again symmetric. We have a closer look at $\widetilde{\mathbold{M}}_j \widetilde{\mathbold{M}}_{j-1}$. The diagonal entries are
\begin{equation}
\begin{aligned}
\tilde c_j \tilde c_{j-1} + \tilde d_j \tilde d_{j-1} &= \frac{1}{4} \left( \frac{\hat m_j }{\hat m_{j+1}} +1 \right) \left( \frac{\hat m _{j-1}}{\hat m_j} +1 \right) \\ 
& \quad + \frac{1}{4} \left( \frac{\hat m_j }{\hat m_{j+1}} -1 \right) \left( \frac{\hat m _{j-1}}{\hat m_j} -1 \right)\\
&=\frac{1}{4} \left( \frac{\hat m_{j-1}}{\hat m_{j+1}} + \frac{\hat m_{j}}{\hat m_{j+1}} + \frac{\hat m_{j-1}}{\hat m_{j}} +1 \right. \\
& \quad \left. + \frac{\hat m_{j-1}}{\hat m_{j+1}} - \frac{\hat m_{j}}{\hat m_{j+1}} - \frac{\hat m_{j-1}}{\hat m_{j}} +1  \right)\\
&= \frac{1}{2} \left( \frac{\hat m_{j-1}}{\hat m_{j+1}} + 1 \right).
\end{aligned}
\end{equation} 

For the off-diagonal entries, we have
\begin{equation}
\begin{aligned}
\tilde c_j \tilde d_{j-1} + \tilde d_j \tilde c_{j-1} &= \frac{1}{4} \left( \frac{\hat m_j }{\hat m_{j+1}} +1 \right) \left( \frac{\hat m _{j-1}}{\hat m_j} -1 \right) \\
& \quad + \frac{1}{4} \left( \frac{\hat m_j }{\hat m_{j+1}} -1 \right) \left( \frac{\hat m _{j-1}}{\hat m_j} +1 \right)\\
&=\frac{1}{4} \left( \frac{\hat m_{j-1}}{\hat m_{j+1}} - \frac{\hat m_{j}}{\hat m_{j+1}} + \frac{\hat m_{j-1}}{\hat m_{j}} -1 \right. \\
& \quad \left. + \frac{\hat m_{j-1}}{\hat m_{j+1}} + \frac{\hat m_{j}}{\hat m_{j+1}} - \frac{\hat m_{j-1}}{\hat m_{j}} -1  \right)\\
&= \frac{1}{2} \left( \frac{\hat m_{j-1}}{\hat m_{j+1}} - 1 \right).
\end{aligned}
\end{equation} 
A simple induction argument shows that the entries of 
\begin{equation}
	\mathbold{\widetilde{M}} = \begin{pmatrix}
	\tilde c & \tilde d\\ \tilde d & \tilde c
\end{pmatrix} \wdef \prod_{j=J}^1 \widetilde{\mathbold{M}}_j 
\end{equation}
are 
\begin{align}
\tilde c &= \frac{1}{2} \left( \frac{\hat m_{1}}{\hat m_{J+1}} + 1 \right)\\
\tilde d &= \frac{1}{2} \left( \frac{\hat m_{1}}{\hat m_{J+1}} - 1 \right).
\end{align}
The relation between the amplitudes is the same as in \eqref{eqn:amplitude_relation}, i.e.
\begin{equation}
\frac{A_{J+1}}{A_1} = \frac{\operatorname{det}(\mathbold{\widetilde{M}})}{ \tilde d} = \frac{m_1}{m_{J+1}} \left( \frac{1}{2} \left( \frac{\hat m_{1} + \hat m_{J+1}}{\hat m_{J+1}} \right) \right)^{-1} = \frac{2 \hat m_1}{\hat m_1 + \hat m_{J+1}},
\end{equation}
which is the same expression as \eqref{eqn:tc_2layer}.

\end{document}